\documentclass[twocolumn, trackchanges, times]{aastex62}
\usepackage{CJK}
\usepackage{amsmath}

\graphicspath{{./}{}}

\received{April 12, 2019}
\revised{xx, 2019}

\submitjournal{ApJ}

\shorttitle{Eclipse Timing and Circumbinary Planets}
\shortauthors{Zhang and Fabrycky}

\begin{document}
\begin{CJK*}{UTF8}{gbsn}
\title{Distinguishing Polar and Co-planar Circumbinary Exoplanets by Eclipsing Timing Variations}
\correspondingauthor{Zhanbo Zhang}
\email{pkuphyzzb@gmail.com}

\author[0000-0003-1411-2663]{Zhanbo Zhang(张湛伯)}
\affil{Institute for Advanced Study, Tsinghua University, Shuangqing Rd.30, Haidian District, Beijing, PRC, 100084 }
\affil{School of Physics, Peking University, Yiheyuan Rd. 5, Haidian District, Beijing, PRC, 100871}

\author[0000-0002-0786-7307]{Daniel C. Fabrycky}
\affil{Department of Astronomy \& Astrophysics, University of Chicago, 5640 S. Ellis Ave., Chicago, IL 60637}

\begin{abstract}

Circumbinary Planets (\textit{CBPs}) can be misaligned with their host binary stars. Orbital dynamics, simulations, and recent observations of proto-planetary disks all suggest that the planet can stably orbit in a plane perpendicular to that of an eccentric host binary star (i.e., a polar orbit). No solid claim of detection of such a configuration has been made; the nine systems detected by the transit technique are nearly coplanar, but their discovery is also biased towards that configuration. Here, we develop Eclipse Timing Variations (\textit{ETVs}) as a method to detect misaligned CBPs. We find that since apsidal motion (periastron precession) of the host binary is prograde for a coplanar planet and retrograde for a polar planet, the mean eclipse period of primary and secondary eclipses differ in a way that distinguishes those configurations. Secondly, the Eclipse Duration Variations (\textit{EDVs}) vary in a way that can confirm that inference, over and against a polar model. Thirdly, the relative phasing of primary and secondary ETVs on the planet's orbital timescale also distinguishes the two configurations, which we explain analytically and quantify through a grid of numerical models. We apply these methods to Kepler-34, a transiting planet known to be nearly coplanar by detailed photodynamic modeling. In this system, we find that the binary eclipse times alone suffice to distinguish these orbital configurations, using the effects introduced here. Our work provides a tool for discovering potential polar CBPs, or misaligned CBPs of milder inclinations, from the existing ETV dataset of the \textit{Kepler}, as well as future observations by \textit{TESS} or \textit{PLATO}.
\end{abstract}

\section{Introduction}
Circumbinary exoplanets, also known as \textit{CBP}s, are one of the most bizarre categories of extrasolar planetary systems, with abundant unsolved questions regarding their formation, migration history, and dynamics. Consistent with the intuitively likely scenario that the circumplanetary disk has the same angular momentum direction as the binary, CBP systems that have been characterized well so far are all nearly coplanar \citep{Welsh12,Doyle11}. However, the coplanarity of CBP systems is not guaranteed theoretically or observationally. At least three pieces of evidence support the existence of misaligned CBPs, as follows.

First, a mutually perpendicular orbital state of CBPs has been found to be static, a state owing its existence to the eccentricity of the binary. The orbital dynamics of such CBP systems was thoroughly studied in \citet{Farago10}, where the authors demonstrated that a polar planetary orbit with its angular momentum aligned with the semi-major axis of the host binary orbit is static. We will call this state \textit{polar} hereafter. \citet{2017Naoz} expanded these studies to an additional order in the expansion of the binary's gravitational potential, and also studied additional apsidal motion of the binary due to general relativity. Simulation work by \citet{Doolin11} found that orbital configurations in which the planet is slightly tilted with respect to the exact polar ones can stay bound, despite undergoing large-amplitude precession. The Kozai-Lidov Effect \citep{Kozai,1962Lidov}, which often trigger instabilities in inclined triple systems, are found by \citet{Martin16} to rarely, if ever, effect the inner binary in CBP systems, due to the low mass of the tertiary. 

Second, assuming the planet formed in a disk of the same inclination as the eventual planet, it is important to know whether disks may be misaligned.  \citet{Martin17}, \citet{Lubow17}, and \citet{Zanazzi17} all determined that a protoplanetary disk around eccentric binaries could warp, then settle, into the into highly-inclined orbital states that \citet{Farago10} identified. The disk requires some initial inclination to settle to the polar state, but a large binary eccentricity allows a wide range of initial conditions to reach that state. Theoretically, these works validate one - but perhaps not the only one - formation scenario of misaligned CBPs. 

Third, thanks to the high-resolution observations of \textit{ALMA}, some misaligned circumbinary disks have been detected, from which misaligned CBPs could possibly emerge via standard disk formation mechanisms \citep{Brinch16,Jensen14,Takakuwa,2019Kennedy}.

Inspired by the pieces of evidence, people have begun working on detecting the misaligned CBPs. An overall review on the kinematic features of misaligned CBPs by \citet{Martin14} pointed out that due to the geometric complexities of such systems, the transiting signals induced by misaligned CBPs would be aperiodic. Transit surveys aimed at finding misaligned CBPs, consequently, have to be conducted on large databases of light curves; the only present survey with long, continuous coverage is the \textit{Kepler} mission \citep{Borucki10}. 

Should we conclude their low detection rate implies their actual low occurrence rate? We should first note that the methodology people have mostly relied on to search for transits is considerably restricted by the low transiting probabilities of such planets, as the tilted orbits have to reside sufficiently far away from the host binary to be stable \citep{Martin15,Martin12,Li16,Doolin11}. 
Therefore, we ask how the indirect tool of Eclipse Timing Variations (hereafter \textit{ETVs}; \citet{Borkovits1, Agol05, Borkovitz2}) can distinguish coplanar from polar orbits. The shape, period, amplitude, and other features in the ETVs contain much dynamical information about the perturber. We aim to show numerically and analytically that we can base inference about planetary orientation on these features.

The study is presented as follows. In section \ref{sec:meth} we briefly describe our adopted numerical methodologies. In section \ref{sec:K34} we describe our attempt to account for the observed ETVs in the system of Kepler-34 with a polar model. Following the failure of the attempt, we show what qualitative differences distinguish coplanar from polar ETV signals in section \ref{sec:phen}. We then move to general discussions of CBP systems. Among the differences, the amplitude ratios and phase shifts of primary and secondary ETVs are particularly diagnostic. We extend our simulations on a much larger grid in the parameter space in section \ref{sec:grid}. The analytic and semi-quantitative explanations to the phenomena are summarized in section \ref{sec:analy}. Finally, we finish by discussing the applicability of our findings to present and future observation data sets and how it could assist in finding more misaligned CBPs in section \ref{sec:diss}.

\section{ETVs of Kepler-34 system}\label{sec:meth}

\begin{deluxetable}{cc}
  \tablecaption{The Kepler-34 system, from \citet{Welsh12}: Orbital parameters in a Jacobian coordinate system, at dynamical epoch $2454969.200$ (BJD).}
  \tablehead{\colhead{Parameter}  &\colhead{Value}}
  \startdata
  \textbf{Planetary Properties} \\
Mass of planet $M_{p}(M_{Jupiter})$ & $0.220^{+0.011}_{-0.010}$ \\
Radius of planet $R_{p}(R_{Jupiter})$ & $0.764^{+0.012}_{-0.014}$ \\
\hline 
\hline
\textbf{Properties of the planetary orbit} \\
Period, $P$ (days) & $288.822^{+0.063}_{-0.081}$ \\
Semi-major axis length, $a$ (AU)& $1.0896^{+0.0009}_{-0.0009}$ \\
Eccentricity, $e$ & $0.182^{+0.016}_{-0.02}$ \\
$e\sin\omega$& $0.025^{+0.007}_{-0.007}$ \\
$e\cos\omega$& $0.180^{+0.016}_{-0.021}$ \\
Mean longitude, $\lambda=M+\omega$ (deg)& $106.5^{+2.5}_{-2.0}$ \\
Inclination, $i$ (deg)& $90.355^{+0.0026}_{-0.0018}$ \\
Relative nodal longitude, $\Omega$ (deg) & $-1.74^{+0.14}_{-0.16}$ \\
\hline
\hline
\textbf{Properties of the stars in the stellar binary}\\
Mass of A, $M_{A}$ ($M_{\odot}$)& $1.0479^{+0.0033}_{-0.0030}$ \\
Mass of B, $M_{B}$ ($M_{\odot}$)& $1.0208^{+0.0022}_{-0.0022}$ \\
\hline
\hline
\textbf{Properties of the stellar binary orbit}\\
Period, $P$ (days)& $27.7958103^{+0.0000016}_{-0.0000015}$ \\
Semi-major axis length, $a$ (AU)& $0.22882^{+0.00019}_{-0.00018}$ \\
Eccentricity, $e$& $0.52087^{+0.00052}_{-0.00055}$ \\
Argument of Periastron, $\omega$ (deg) &$71.44\pm0.06$ \\
$e\sin\omega$& $0.49377^{+0.00057}_{-0.00060}$ \\
$e\cos\omega$& $0.165828^{+0.000065}_{-0.000061}$ \\
Mean longitude, $\lambda=M+\omega$ (deg)& $300.1970^{+0.0099}_{-0.0105}$ \\
Inclination, $i$(deg)& $89.8584^{+0.0099}_{-0.0105}$ \\
\hline\hline
  \enddata
\end{deluxetable}\label{tab:K34params}

One of the few known CBPs with well-observed ETVs is Kepler-34 \citep{Welsh12}. It also transits, and the lightcurve has been modeled, showing it to be on a nearly coplanar orbit. In this section, we aim to show that the ETVs alone can yield that inference. Hence this test-case will lead us to develop the properties that a general system could reveal its orientation.

\subsection{Why Kepler-34 System and Static Polar States?}
Kepler-34 system is reported in \citet{Welsh12} utilizing only data from the first 671 days of \textit{Kepler} Mission. The orbital parameters of the binary and the planetary companion are all solved to considerable accuracy by analyzing the Radial Velocity (RV) and light curves, as listed in Table \ref{tab:K34params}. Bearing two solar-mass stars and a sub-Saturn co-planar planetary orbit, the system has the following advantages that made it a suitable template of our study:

\begin{enumerate}
    \item The binary eccentricity $e_1=0.52$ is the highest among known CBP systems. Since the binary eccentricity has been found to be an inclination-stablizing mechanism, Kepler-34 might be akin to binary systems hosting polar CBPs.
    \item With a moderate planetary period and bright luminosity, the system allows for precise and long-term ETV measurements. Up to now, about 50 primary eclipses and secondary eclipses have been recorded with their timings measured precisely. We present the ETVs in figure \ref{fig:K34} from the data through the end of the mission (see appendix).
    \item The primary and secondary masses in the binary system are almost the same. Thus, the primary eclipses and secondary eclipses are of similar depths, consequently giving rise to similar uncertainties on their timings ($\sim 2s$ and $\sim4s$ respectively). This helps traits concerning both primary and secondary ETVs stand out.
\end{enumerate}

 \begin{figure*}[!thb]
  \centering
   \includegraphics[width=\linewidth]{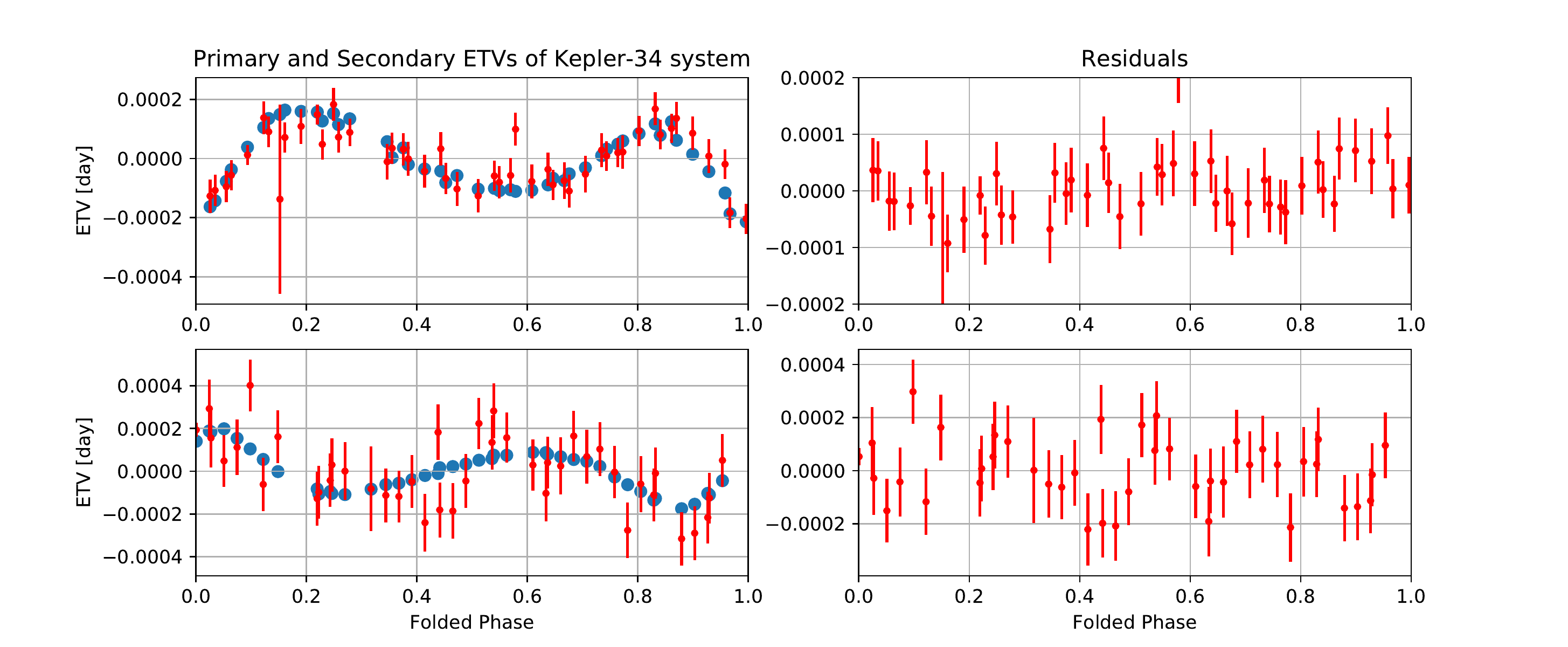}
  \caption{Primary and secondary ETVs of the Kepler-34 system folded into a single period. The upper band is for primary, the lower is for secondary. Blue dots are from our numerical simulation of the model described by parameters in table \ref{tab:K34params}, and red points with error bars are from \textit{Kepler} observations. The right column is for the residuals. Uncertainties are rescaled to make the standard deviation of residuals close to 1.}
  \label{fig:K34}
\end{figure*}

\begin{figure*}[!thb]
  \centering
   \includegraphics[width=\linewidth]{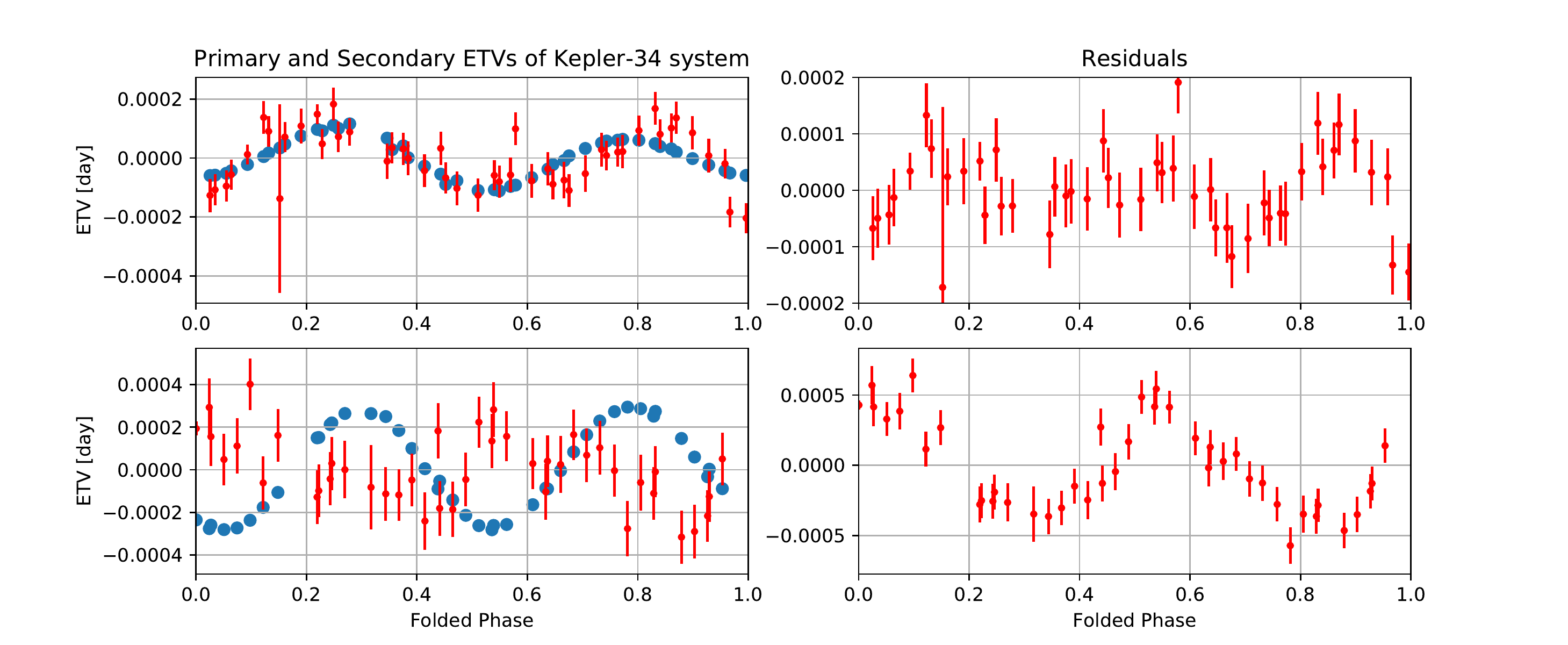}
  \caption{ETVs produced by the best fitting polar model vs observation values. Uncertainties are the same as in figure \ref{fig:K34}.}
  \label{fig:PolarK34V}
\end{figure*}

Rather than exploring the whole parameter space of the potential misaligned companion, we focused on the static polar ones. The analytic work in \citet{Farago10} and numerical simulations in \citet{Doolin11} revealed that the polar states are the only other category corresponding to a dynamical fixed-point, besides the coplanar ones. Systems moderately misaligned from other of these states simply precess around them, and they should yield similar ETV signatures. Octopole effects can chaotically transition orbits between these two precession states, but only for large-angle librations \citep{2017Naoz}, which are not likely to persist in dissipative disks \citep{Martin17,Zanazzi17}.

A minor point needs to be clarified. Static polar orbits, by definition, actually contain two sub-categories with their inclination equals $90^{\circ}$ or $-90^{\circ}$, respectively. The symmetry indicates that the ETV features of them would be of no real difference, as will be confirmed in section \ref{sec:analy}, where the mutual inclination between the binary and the planetary orbit $i$ are involved in the analytic expressions in the form of $\sin^2 i$ or $\cos i$. Thus in the following texts, we would be focused only on those orbits with $i=90^\circ$.

\subsection{N-body Simulation and ETV Extraction}
We next present how we derive the eclipsing timings with N-body simulations using the \texttt{python} package \texttt{rebound} \citep{Rein12}.

All the simulations were conducted with integrator \texttt{ias15} \citep{IAS15}. Jacobian coordinates were adopted \citep{Fabrycky10}. We take the conventional coordinate system in which the observer watches the system from $-z$ direction, and thus, an edge-on binary system with ascending node $\Omega=0$ lies in the $xOz$ plane. 

We trace the mutual eclipsing events of the simulated particles in a way illustrated in \citet{Fabrycky10}: the projectile distance $\mathbf{r}_{1,2} = \mathbf{r}_1-\mathbf{r}_2 $ of the two bodies of interest on the sky-plane (namely the $xOy$ plane) is followed. When the projectile distance $|\mathbf{r}_{1,2}|$ reaches its minimum, we have the following function vanish:

\begin{equation}\label{eq:project}
    g(x_{1,2},y_{1,2},\dot{x}_{1,2},\dot{y}_{1,2}) = \mathbf{r}_{1,2} \cdot \dot{\mathbf{r}}_{1,2} =x_{1,2}\dot{x}_{1,2}+y_{1,2}\dot{y}_{1,2}
\end{equation}

We divide the time-span of a simulation process into 4000 pieces, and stop by whenever $g$ changes its sign. Within the sign changing range, we conduct a root-finding program using the bisect algorithm embedded in \texttt{scipy} \citep{scipy} to locate the closest approach. If the relative distance at the closest approach is indeed less than the sum of the radii of the two bodies, a transit/eclipse is recorded. 

We noted that the reduced $\chi^2$ of the best orbit model from \citet{Welsh12} to the observed eclipsing timings are $\sim$ 11.3 and $\sim$ 9 in the primary and secondary cases respectively, whereas they should be $\sim 1$. We attribute this to an underestimation on the eclipsing timing uncertainties, and we multiply them by an overall factor, to enforce reduced $\chi^2=1$. We have included eclipses affected by outlying photometric points, planetary transits, and incomplete accounting for slopes and curvature that are either instrumental or astrophysical. It would be possible to make a cleaner model with smaller timing residuals, by strictly choosing the eclipses that can be measured cleanly and excluding others. 

We then decompose the timings with the following equation by doing an optimization of the residuals from the true values to the model using \textit{Levenberg-Marquardt} algorithm implemented by \texttt{scipy}:

\begin{equation}\label{eq:decomp}
    T_{i}=T_{0}+P\times i+\sum_{j=1}^{3}(A_{j}cos(2\pi f t)+B_{j}sin(2\pi f t))
\end{equation}

In which $P$ denotes the binary period, and $f$ denotes the base frequency of the ETVs, which in most cases of interest, is approximately the inverse of the perturber's period. Subtracting the first two terms from the eclipsing times can we obtain the ETVs. We chose to include only three Fourier harmonic terms in that with the available measurement accuracy higher order terms cannot be accurately extracted. 

In this way we decompose the observed ETVs of Kepler-34 system, and list the results in table \ref{tab:Fourier}. One can find the exact values of the eclipsing timings in the appendix (see table \ref{tab:PETVs} and \ref{tab:SETVs}  ).

\begin{deluxetable}{lrr}
  \tablecaption{Fourier parameters of ETVs in the Kepler-34 system}
  \tablehead{\colhead{Parameter}&\colhead{Primary}&
  \colhead{Secondary}}
  \startdata
$T_{0}$, JD-2454900 & 51.982079(7) & 69.179925(7)\\
Period P (days)& $27.7954091(2)$ & $27.7953702(2)$ \\
Frequency f (1/days) &0.003515(5)& 0.003513(7)\\
$A_{1}$ (days)& $1.98(50)\times10^{-5}$ & $-1.02(51)\times10^{-5}$ \\
$B_{1}$ (days) & $5.24(49)\times10^{-5}$ & $-7.15(38)\times10^{-5}$\\
$A_{2}$ (days)& $-1.05(5)\times10^{-4}$ & $7.19(76)\times10^{-5}$\\
$B_{2}$ (days)& $-3.53(79)\times10^{-5}$ & $8.89(68)\times10^{-5}$\\
$A_{3}$ (days)& $-6.16(54)\times10^{-5}$ & $4.41(66)\times10^{-5}$\\
$B_{3}$ (days)& $-1.89(69)\times10^{-5}$ & $4.54(66)\times10^{-5}$\\
\hline
\enddata
\end{deluxetable}\label{tab:Fourier}

\begin{deluxetable}{lrr}
  \tablecaption{Fourier parameters of ETVs produced by a polar model that best accounts for the primary ETVs }
  \tablehead{\colhead{Parameter}  &\colhead{Primary} &\colhead{Secondary}}
  \startdata
$T_{0}$, JD-2454900 & 51.928312(1) & 69.179617(5)\\
Period P (days)& $27.79523419(3)$ & $27.79587800(2)$ \\
Frequency f (1/days) &0.003505(1) & 0.0035046(2) \\
$A_{1}$ (days) & $2.02(7)\times10^{-5}$ & $-0.82(3)\times10^{-5}$ \\
$B_{1}$ (days)& $2.37(6)\times10^{-5}$ & $-1.76(4)\times10^{-5}$\\
$A_{2}$ (days)& $-7.96(8)\times10^{-5}$ & $-25.02(5)\times10^{-5}$\\
$B_{2}$ (days)& $-3.07(13)\times10^{-5}$ & $-13.29(3)\times10^{-5}$\\
$A_{3}$ (days)& $0.26(3)\times10^{-5}$ & $0.97(3)\times10^{-5}$\\
$B_{3}$ (days)& $-0.07(2)\times10^{-5}$ & $-0.27(2)\times10^{-5}$\\
\hline
\enddata
\end{deluxetable}\label{tab:PolarFourier}

\subsection{A Polar Kepler-34 b?}

\subsubsection{Dynamical Fitting}\label{sec:K34}
With the methodologies developed in the last chapter, we then attempt to fit the observed ETVs in Kepler-34 system with a planetary companion residing on a static polar orbit. As was pointed out by \citet{Farago10}, the orbital plane of such orbits is entirely certain given the binary orbit. Therefore, there are five free parameters remaining in our polar model.

\begin{enumerate}
    \item Planetary mass $m_C$. 
    \item Planetary semi-major axis $a_2$, or equivalently, the planetary period $P_2$. The ETV period is dependent on it, thus, regardless of the orbital configuration, we could expect it to be a similar value compared with the co-planar model. Hence, in the fitting process, we could put in a well-constrained prior distribution of it.
    \item Planetary Eccentricity $e_2$.
    \item Initial Mean anomaly $M_2$ , which sets the initial phase of the ETVs.
    \item Periastron angle $\omega_2$.
\end{enumerate}

We used Markov Chain Monte Carlo(\textit{MCMC}) to fit the observed eclipsing timings in Kepler-34 system with a static polar planetary perturber. The implementation we used is the \texttt{python} package \texttt{emcee} \citep{emcee}. The integrator simulates the \textit{Kepler} mission time span of 4 years, yielding the eclipsing timings from a given model, which are then compared with the observed values (tables \ref{tab:PETVs} and \ref{tab:SETVs}) to construct a posterior distribution via MCMC. 

\begin{figure}[!thb]
  \centering
   \includegraphics[width=\linewidth]{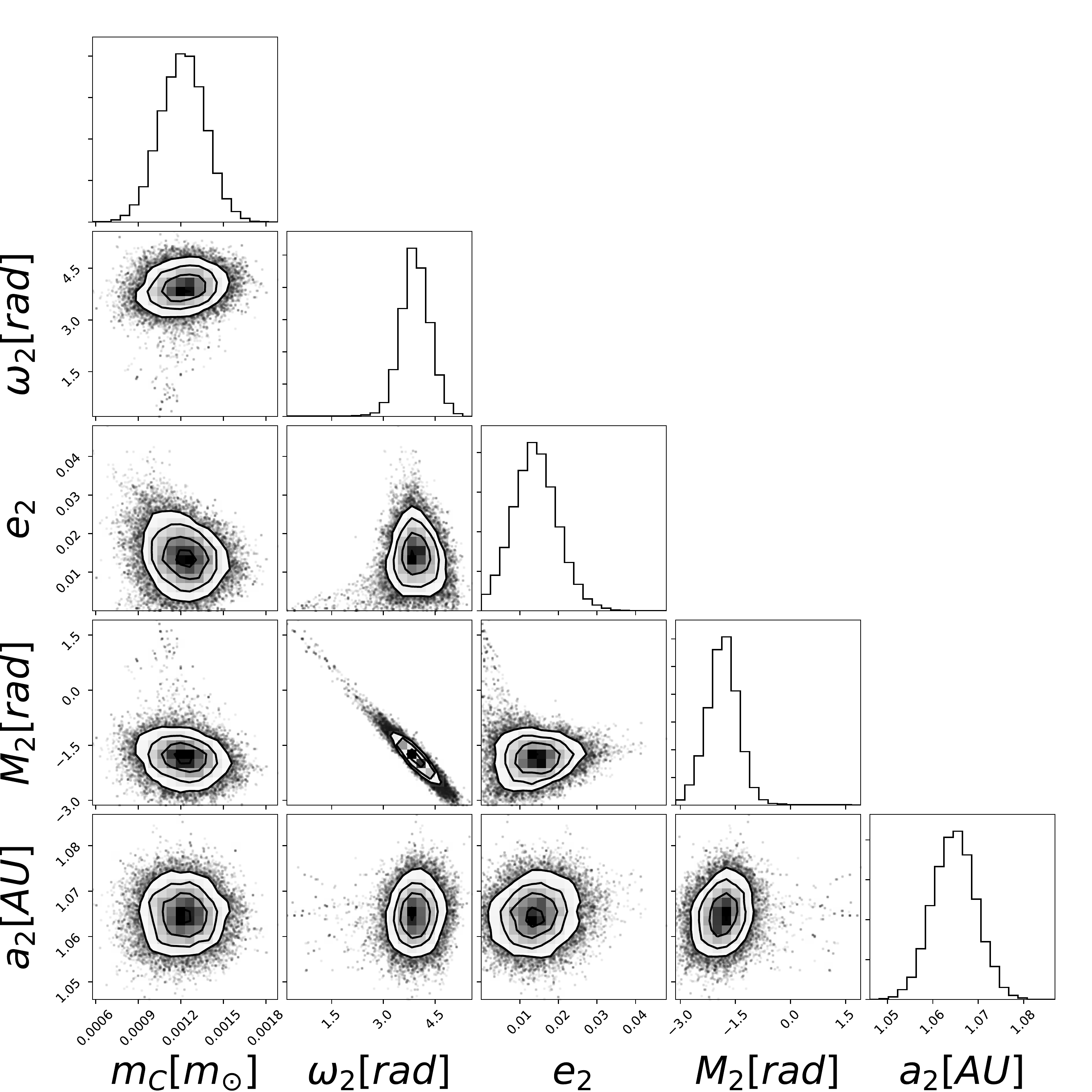}
  \caption{Corner plot for the MCMC fitting of polar model to Kepler-34 primary ETVs. Degeneracy of $M_2$ and $\omega_2$ are due to the fact that the mean longitude $\lambda=M+\omega$ together decides the initial phase of the planetary orbit.}
  \label{fig:CornerK34}
\end{figure}

We first included only the primary eclipses, when the fitting yields a best fitting model with a reduced $\chi^2_{43}\approx1.6$. The best fitting parameters of the particular model are listed in table \ref{tab:PolarK34}, and a corner plot for the MCMC is available in figure \ref{fig:CornerK34}. 

On the other hand, when the primary eclipsing timings are fitted with a co-planar planet, we acquired a result with a reduced $\chi^2_{43} \sim 0.75$. Note this model is even better than the model solved in \citet{Welsh12}, which has to account for the secondary eclipses simultaneously. Comparing the exact polar model and exact co-planar model fitting the primary eclipsing timings, we shall say that the reduced $\chi^2_{43} \sim 1.6$ is not preferred, yet not large enough to be completely ruled out. From figure \ref{fig:PolarK34V}, we could see that the polar model can capture the overall shape of the primary ETVs, and its larger residuals could indicate its failure depicting some detailed features, as well as a slight difference in the periods (see section \ref{sec:phen} for more discussions). 

\begin{deluxetable}{lll}
  \tablecaption{Best-fitting parameters in a polar model for Kepler-34's primary ETVs}
  \tablehead{\colhead{Parameter}  &\colhead{Unit} &\colhead{Value}}
  \startdata
$M_{C}$ & $M_{J}$ & $1.28^{+0.17}_{-0.17}$ \\
$\omega_{2}$ & deg & $225^{+17}_{-16}$ \\
$e_2$ && $0.014^{+0.006}_{-0.005}$ \\
$M_2$ &deg & $-106^{+22}_{-23}$ \\
$a_2$ & AU & $1.065^{0.005}_{0.005}$ \\\hline
\enddata
\end{deluxetable}\label{tab:PolarK34}

We then considered the secondary eclipses. We first applied the polar model that could fairly account for the primary ETVs to the secondary ones, as is shown in figure ~\ref{fig:PolarK34V}. It is obvious that the phase of the observed secondary ETVs is anti-aligned with those produced by the polar model. In addition, modeling both the primary and secondary ETVs as the target data with a polar model leads to an unacceptably large reduced $\chi^2 \sim5$.

Therefore, we came to the conclusion that in the specific case of Kepler-34 system, the primary ETV signals could be accounted for with either a coplanar or a polar CBP perturber, although the polar perturber is statistically not preferred,  while when considering also the secondary ETVs, the polar model fails to work. 

\subsubsection{Phenomenology}\label{sec:phen}
The preceding subsection has revealed that the joint features of the primary and secondary ETVs might exclusively indicate the geometrical configuration of the planetary perturber. In this subsection, we present what the differences are in the ETV signals induced by a co-planar and a polar planet that make them distinct.

For the convenience of comparison, we decomposed the ETVs produced by the best fitting polar model in the manner described in section \ref{sec:meth}, and we tabulate the results in table \ref{tab:PolarFourier}.

\textbf{1. Primary and Secondary Eclipsing Periods Discrepancy}\par
The second lines in both table \ref{tab:Fourier} and table \ref{tab:PolarFourier} show that the primary and secondary eclipses have their periods differing by a tiny, yet detectable amount. We will revisit this quantitatively in section \ref{sec:precession}.  

The primary and secondary periods measured in Kepler-34 system are $P_{pri}=27.7958063(16)$~days and $P_{sec}=27.7957548(17)$~days, respectively, giving rise to a discrepancy of $4.4\times10^{-5}$~day, or rather, about 4 seconds\footnote{The periods measured here are different than the values present in table 
\ref{tab:Fourier}, which is due to the uncertainty in the binary semi-major axis as an input orbital parameter rather than the periods per se. }. On the other hand, the period discrepancy from the polar model that could explain the primary ETVs is as large as 54 seconds, much larger than the observed value, and with a different sign. 

\textbf{2. The Eclipsing Duration Variations (\textit{EDVs})}\par
\begin{figure}[!thb]
  \centering
   \includegraphics[width=\linewidth]{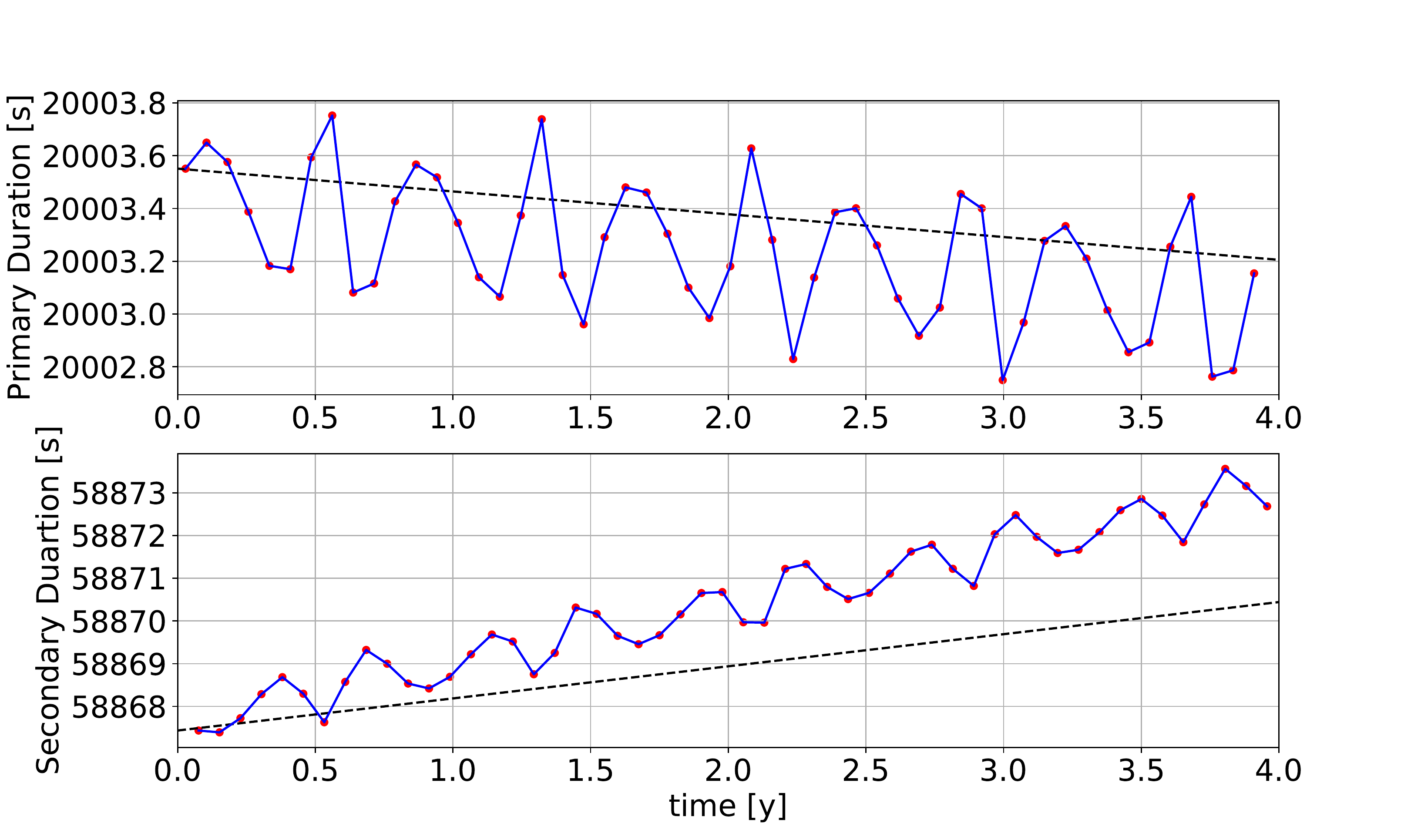}
  \caption{Simulated EDVs in the best fitting coplanar Kepler-34 system,\textit{top}: primary, \textit{bottom}: secondary. The black dashed lines are predicted by equation \ref{eq:EDVs}.}
  \label{fig:CopEDV}
\end{figure}
\begin{figure}[!thb]
  \centering
   \includegraphics[width=\linewidth]{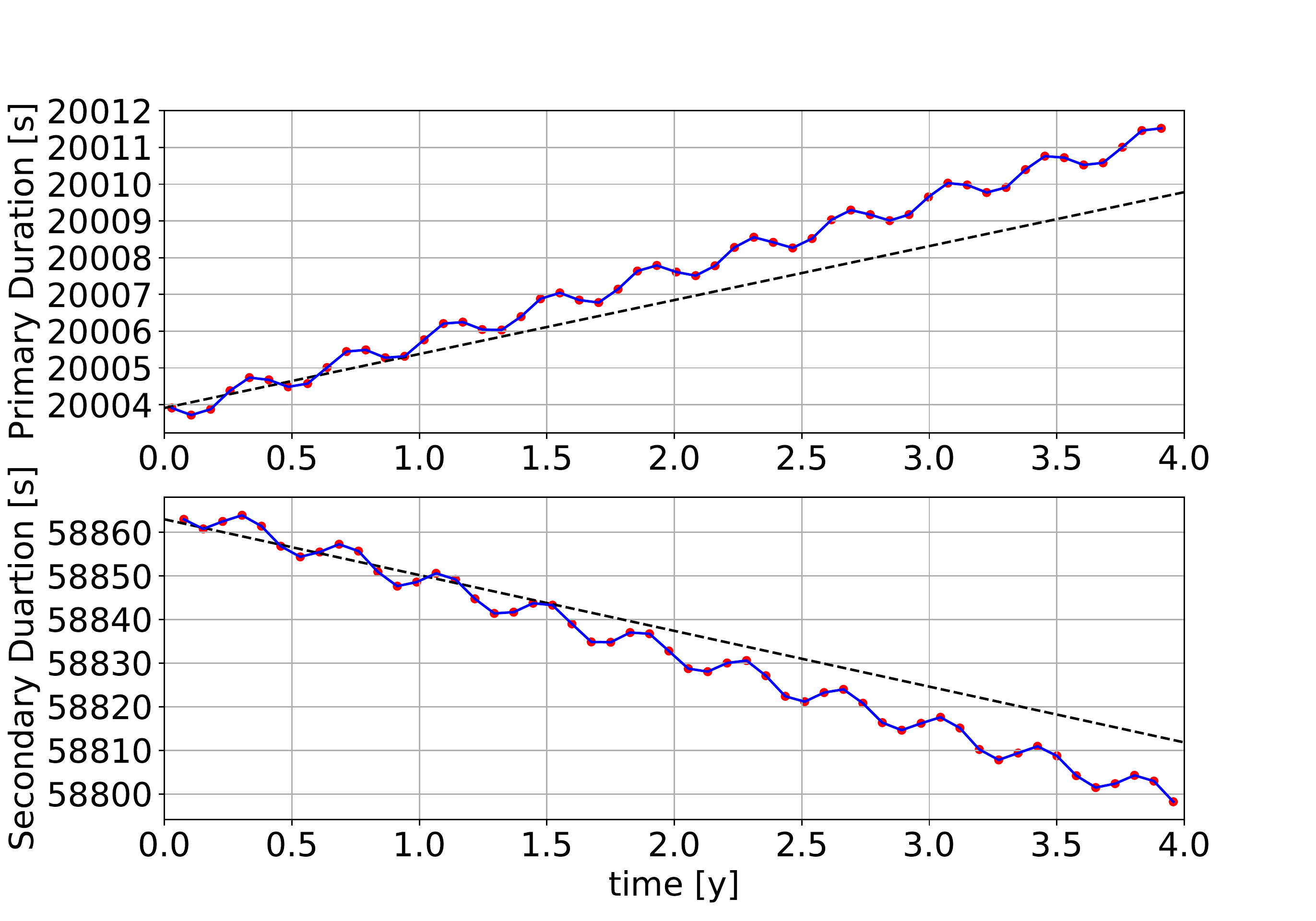}
  \caption{Simulated EDV in the best fitting polar model of Kepler-34 system. The black dashed lines are predicted by equation \ref{eq:EDVs}.}
  \label{fig:PolEDV}
\end{figure}

Similar to the period discrepancy, eclipsing duration variations, reflecting the evolution of the orientation of the binary orbit towards the observer induced by precession, could be of considerably different
magnitudes, and opposite sign in the two plausible configurations. To demonstrate that, we plot the EDVs from our simulation of Kepler-34 with the model in \citet{Welsh12} and the polar model in figures \ref{fig:CopEDV} and \ref{fig:PolEDV}, respectively.

In addition to a short-term variation induced by the orbital motion of the planet, the long-term variations in EDVs induced by the \textit{apse-nodal} (a term from \citet{Borkovitz2}) precession of the binary orbit shown in the two figures are completely different. First, their amplitudes differ by a factor of about $17$, the same as the proportion in the period discrepancy, which hints that it is the same mechanism that underlies these two phenomena. Actually, the long-term EDV trend in the coplanar case is so small (less than 5 seconds over the time span of 4 years) that measuring it with observational data may not be probable, given the accuracy in eclipsing duration measurement is comparable to that in the eclipsing timing measurements.

Similarly, the opposite precession directions seem to cause the EDVs in the two cases to evolve in different directions -- in the co-planar case, the primary eclipsing durations decrease, while the secondary ones increase, in the polar case, vice versa. 

\textbf{3. Apse-Nodal Precession Direction and Speed}\par
\begin{figure*}[!thb]
  \centering
   \includegraphics[width=\linewidth]{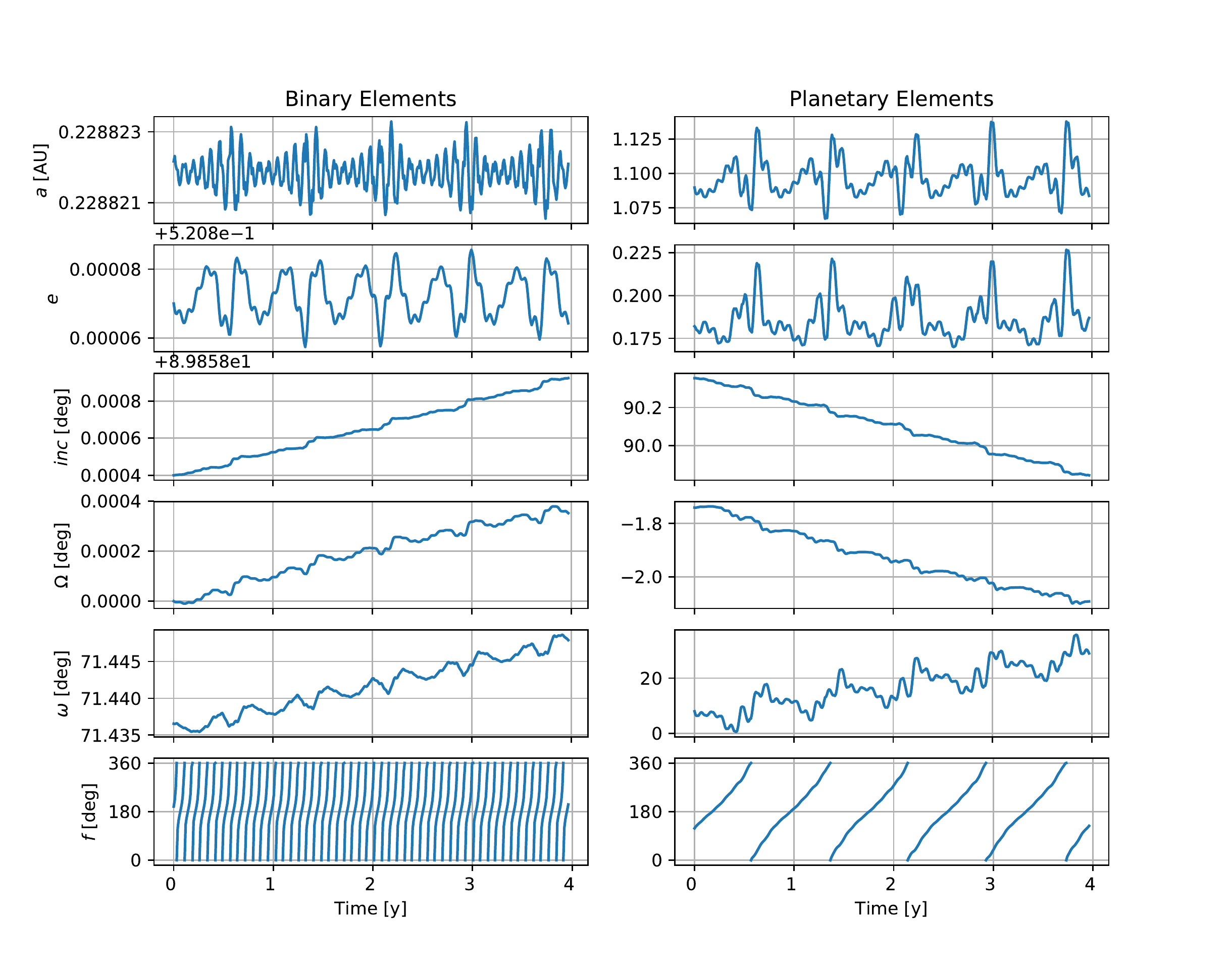}
  \caption{Orbital elements variation over the timespan of 4 years for the actual Kepler-34 system. The left column is for the binary orbital elements, right for the planetary ones. $a$: Semi-major axis length; $e$: Eccentricity; $inc$: Orbital Inclination in the Observers' Frame; $\Omega$: Nodal Longitude; $\omega$: Periastron Angle; $f$: True Anomaly. All variables are Jacobian. The half-planet period oscillation of all variables are due to the perturbation of the planet. As in the actual system the binary is slightly tilted from the $xOz$ plane, the long term precession also drives $\Omega$ and $i$ to change in the long run. }
  \label{fig:CopElements}
\end{figure*}
\begin{figure*}[!thb]
  \centering
   \includegraphics[width=\linewidth]{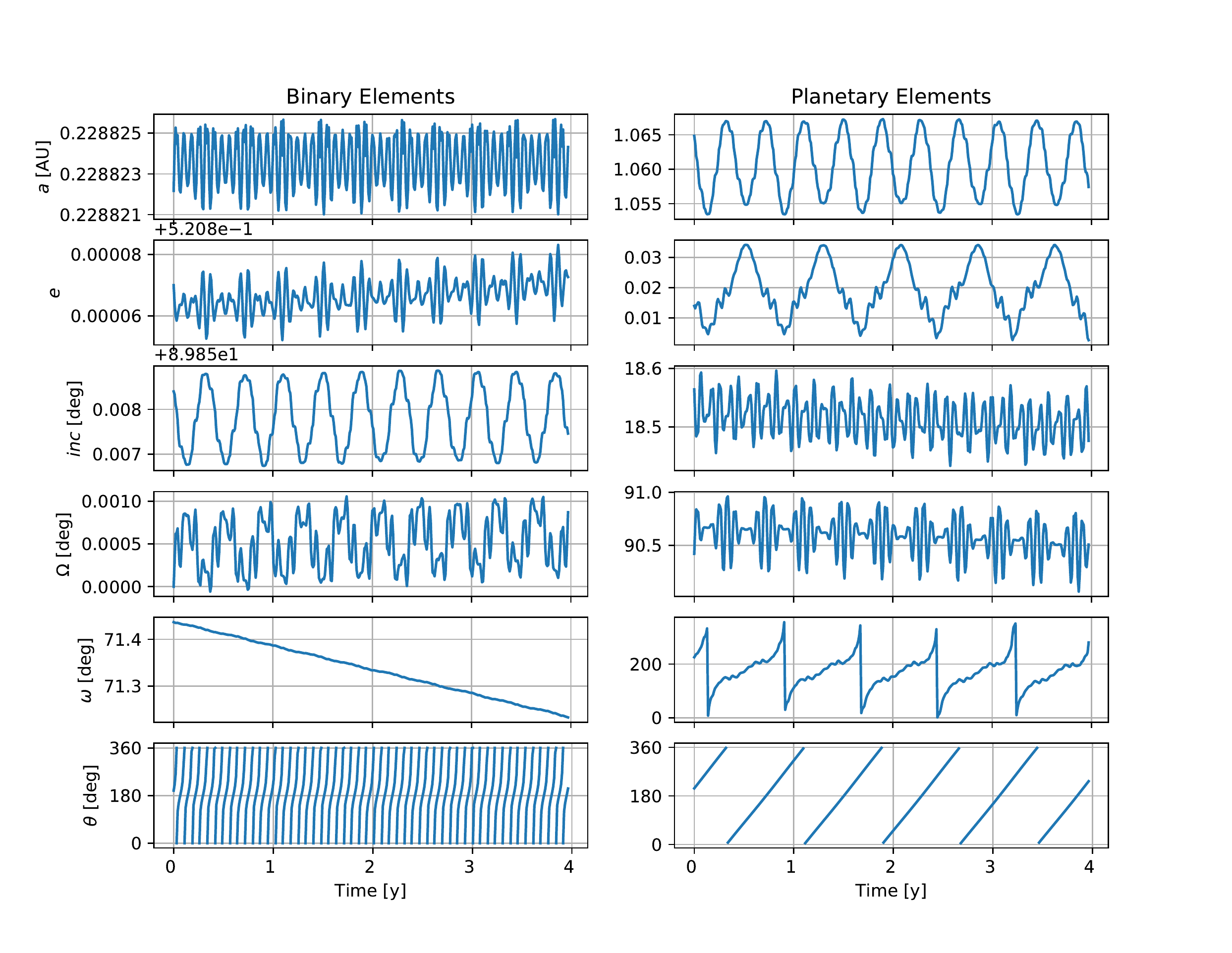}
  \caption{Orbital elements variation over the timespan of 4 years for the best fitting polar model. The left column is for the binary orbital elements, right for the planetary ones. $a$: Semi-major axis length; $e$: Eccentricity; $inc$: Orbital Inclination in the Observers' Frame; $\Omega$: Nodal Longitude; $\omega$: Periastron Angle; $\theta$: True Anomaly. Note for the planet we have used $\theta_2 = \arctan(y_2/\sqrt{x_2^2+z_2^2})$ instead of the output from \texttt{rebound}, as the latter one is measured with respect to $xOz$ plane rather than in the orbital plane of the planet itself. Other parameters are Jacobian.  The half-planet period oscillation of all variables are due to the perturbation of the planet. As in the actual system the binary is slightly tilted from the $xOz$ plane, the long term precession also drives $\Omega$ and $i$ to change in the long run.}
  \label{fig:PolElements}
\end{figure*}

Prompted by the two foregoing phenomena, we also plot the variation of the 12 orbital parameters in the two cases over the timespan of the simulation, as is shown in figures \ref{fig:CopElements} and \ref{fig:PolElements}. The different direction and speed at which the binary periastron orientation $\omega_1$ precesses are conspicuous. If the radial velocity of a system has been observed over many years, this effect can show up in the drift of $\omega$ \citep{2000Jha}. 

\textbf{4. Morphological Distinctions of the Coplanar and Polar ETV Curves}\par
Another difference of the ETVs in the two configurations are the phase shift and the amplitude ratio between the largest harmonic component of the primary and secondary ETV curves. To illustrate them, we define the following two quantities as the `\textit{strength}' and `\textit{relative phase}' of the $i$th frequency component in the Fourier expansions \ref{eq:decomp}.

\begin{equation}
\label{eq:Am}
F_{i}=\sqrt{A_{i}^{2}+B_{i}^2}
\end{equation}

\begin{equation}
\label{eq:phi}
\phi_{i}=arctan(A_{i}/B_{i})+sgn(B_{i})\pi
\end{equation}

We present values of these quantities in the observed ETVs of Kepler-34 system, and in the ETVs produced with the polar model in table \ref{tab:Shape}.

\begin{deluxetable*}{lllll}
  \tablecaption{Parameters Describing Shapes of ETVs in Polar and Coplanar Model in Kepler-34 system }
  \tablehead{\colhead{Value}  &\colhead{Primary} &\colhead{Secondary} & \colhead{Amplitudes Ratio} &\colhead{Phase Shifts} \\ &&&\colhead{$R_{i}=F_{pi}/F_{si}$}&\colhead{$S_{i}=0.5-|(\phi_{pi}-\phi_{si})mod(2\pi)-0.5|$}\tablenotemark{a} }
  \startdata
\textbf{Coplanar} \\
\hline\hline
$F_{1}$ & 5.6e-5 & 7.2e-5 & 0.776 & Not Applicable(N/A) \\
$\phi_{1}$, rad & 0.361 & 3.28  &N/A&  0.536 \\
$F_{2}$ & 11.07e-5 & 11.45e-5 & 0.967 & N/A \\
$\phi_{2}$, rad & 4.39 & 0.68 &N/A& 0.590 \\
$F_{3}$ & 6.44e-5 & 6.32e-5 & 1.02 & N/A \\
$\phi_{3}$, rad & 4.41 & 0.77 & N/A & 0.576 \\
\hline\hline
\textbf{Polar} \\
\hline\hline
$F_{1}$ & 3.11e-5 & 1.94e-5 & 1.60 & N/A  \\
$\phi_{1}$, rad & 0.705 & 3.58  &N/A&  0.542 \\
$F_{2}$ & 8.53e-5 & 28.3e-5 & 0.301 & N/A \\
$\phi_{2}$, rad & 4.34 & 4.66 &N/A& -0.05 \\
$F_{3}$ & 0.27e-5 & 1.00e-5 & 0.27 & N/A \\
$\phi_{3}$, rad & 1.83 & 1.84 & N/A & 0.00 \\
\hline\hline
\enddata
\tablenotetext{a}{By this definition, $S_i$ would be close to 0 when the two phases are aligned, and close to 0.5 when they are shifted by $\pi$.}
\end{deluxetable*}\label{tab:Shape}

The second-order terms in the Fourier expansions dominate in both cases, hence we would focus on the amplitude ratio $R_2$ and phase shift $S_{2}$. In the co-planar case, the primary and secondary ETVs are of similar amplitudes, yet their phases differ by $\sim\pi$, while when the perturber is on a polar orbit, the $R_2$ goes conspicuously below 1, and the primary and secondary ETV phases align. 

To sum up, we have noticed that in the case of Kepler-34, there are various aspects in the properties of ETVs induced by a polar or a co-planar planetary companion that are distinct, and thusly the observed ETVs of Kepler-34 system could not be induced by a planet residing on the static polar orbit. In other words, the ETVs alone are capable of ruling out the polar model as a feasible solution to the system. 

\section{ETVs in General CBP Systems}
\subsection{Grid Simulations}\label{sec:grid}

In the previous section, it was noted from the dynamical fitting that the amplitude ratio, as well as the phase shift between primary and secondary ETVs, are effective indicators of the geometrical configuration of the companion's orbit in Kepler-34 system. In this section, we will extend the coverage of them as indicators of the geometric configuration of the system to wider parts of the parameter space of binary hosts and planetary parameters, via N-body simulations on a grid of the parameter space. 

We will consider the influence of these orbital parameters of the CBP system:

\begin{enumerate}
    \item The binary mass ratio $q=\frac{M_2}{M_1}$. We will fix $M_1$ to 1 solar mass and alter $M_2$ in range $(0.08,1)M_\odot$ to control the parameter. The radii of the stars needed for transit/eclipse searching in the simulations are estimated from the mass according to \citet{Demircan91}, but this only effects the EDVs, not the other phenomena. 
    \item The geometrical parameters of the binary system: eccentricity $e_1$ and periastron angle $\omega_1$
    \item The period ratio of the outer and inner orbits $\frac{P_2}{P_1}$. Similar to the mass ratio, we will keep the binary period fixed to be $27.79$ days as is in Kepler-34 system, and let $P_2$ vary. 
    \item The geometrical parameters of the companion, namely $e_2$ and $\omega_2$. 
\end{enumerate}

The only remaining free parameter in a CBP system is the planetary mass, which simply decides the amplitude of the ETVs, and would not result in any changes in the morphological behaviors of the ETVs. We, therefore, fix it to be $0.22M_J$, as is in Kepler-34 system.

Even specializing to circular orbits, the parameter space is four-dimensional. It would be not only unacceptably computationally expensive but also impossible to visualize if we explore the four parameters altogether. Instead, we run our simulations on several grid on combinations of two of the parameters. On each grid point, a co-planar and a polar CBP is generated with the same mass on a circular (except when we focus on $e_2$) orbit. We set on each grid point two random initial phases for the primary and secondary orbits. The simulations run for 5 planetary periods. The ETVs were then recorded, extracted, decomposed, and analyzed in a pipeline. We focus on the second term in the Fourier expansions of the ETVs, as it is almost always the dominant term. Then, four color images in which the amplitude ratios and phase shifts are drawn. There are combinations of input orbital parameters that would lead to unstable planetary orbits out of the stability zone of \citet{Holman99}, from which no ETVs could be drawn. We left them blank in the images. The results are as below.

\subsubsection{The Geometrical Parameters of the Host Binary}

We fixed $P_2=300$ days, $M_2=0.5M_{\odot}$ and $e_2=0$. The grid on which we ran the simulation are $e_1$ in $[0,0.05,0.1,...0.95]$ and $\omega_1$ in $[0,\pi/6,\pi/3,...2\pi]$.

\begin{figure*}[!thb]
  \centering
   \includegraphics[width=\linewidth]{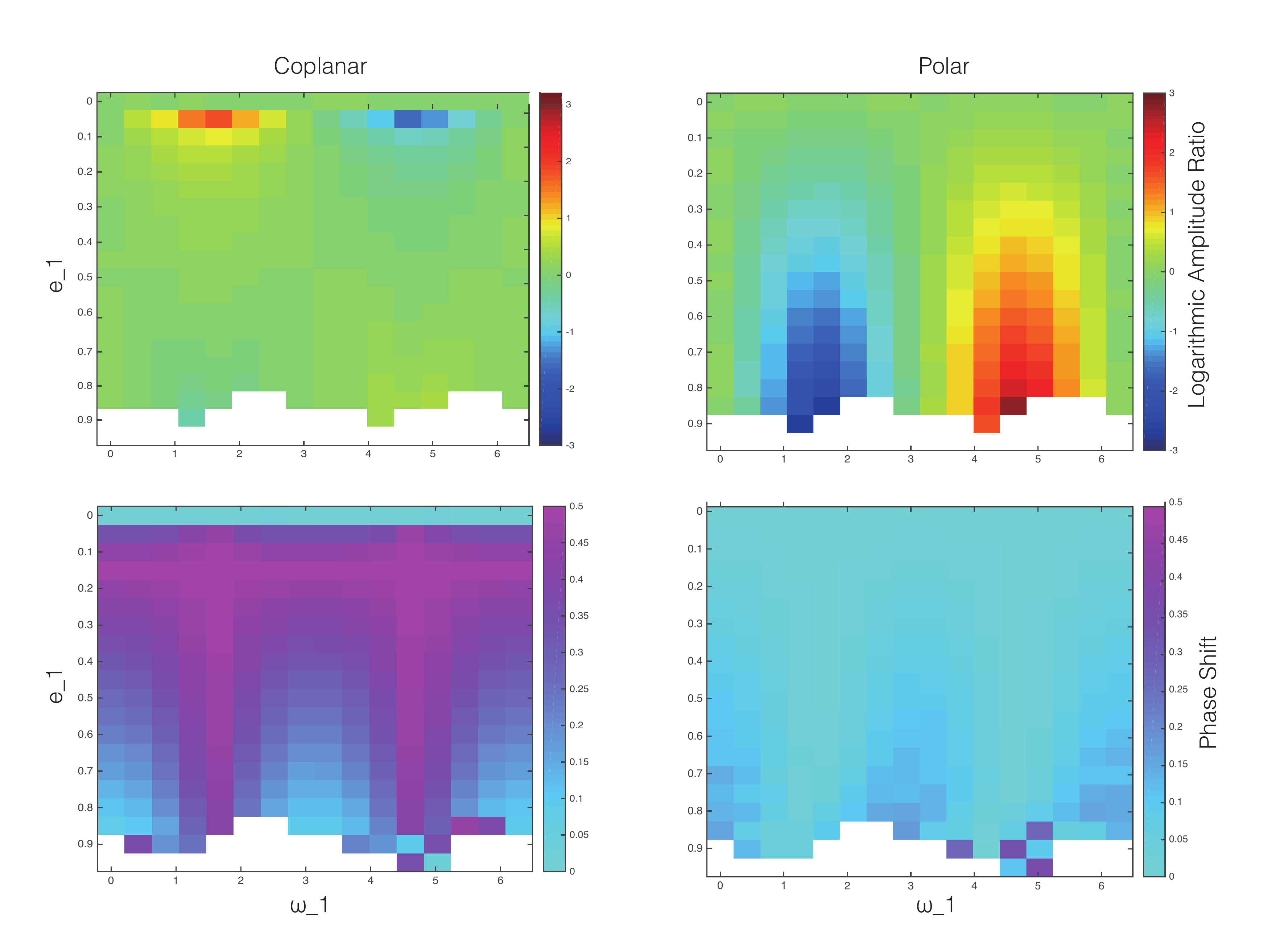}
  \caption{The maps of amplitude ratio and phase shift on the $e_1-\omega_1$ grid. The left column is for the co-planar case, while on the right are from ETVs induced by a polar planet. The top panel and bottom panel denote the amplitude ratio and phase shifts respectively. The color maps of amplitude ratios and phase shifts visualize $ln(R_{2})$ ($e$ based), and $S_2$ respectively. $S_2$ and $R_2$ are defined in table \ref{tab:Shape}.}
  \label{fig:eomesh}
\end{figure*}

The results are visualized in figure \ref{fig:eomesh}. It is clear that the maps of amplitude ratio and phase shifts over the $e_1-\omega_1$ grid are significantly different for the two categories of CBPs. Despite small zone when $e_1$ is negligible, the phase shift is mostly misaligned for the coplanar planet, while the whole phase shift map of the polar planet shows alignment. In the amplitude ratio maps, the co-planar planet generally induces primary and secondary ETVs of similar amplitudes, but abnormalities occur when $\omega_1=0.5\pi$ or $1.5\pi$, and $e_1$ being sufficiently small but not zero, where the amplitude ratio drops far below or far above 1. On the contrary, the amplitude ratio behaves regularly at small inner eccentricities, but seems to rise up when the binary orbit becomes more eccentric in the polar case.

There is a sudden change in both amplitude ratio and phase shifts in the co-planar case when $e_1$ is adequately small. Therefore, we ran a second set of simulations on finer grid where the $e_1$ grid is $[0,0.005,0.01,...0.1]$. As is shown in figure 9, the sudden change on the crude grid is actually continuous, with a peak and a valley of the amplitude ratio occurring at $e_1\sim0.03$.

\begin{figure}[!thb]
  \centering
   \includegraphics[width=\linewidth]{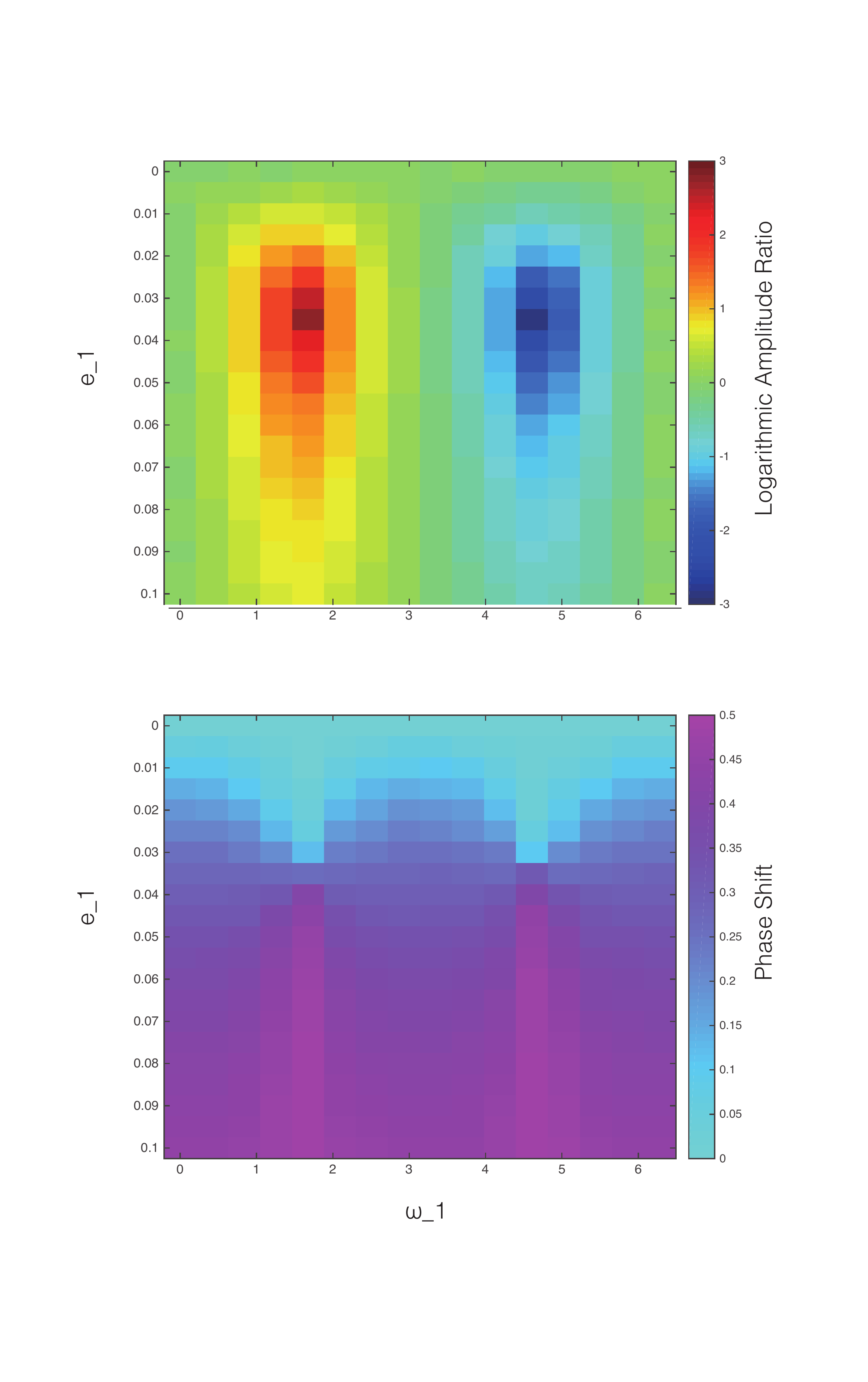}
  \caption{The maps of amplitude ratio and phase shift on the $e_1-\omega_1$ mesh grid in which $e_1$ is resolved on a finer basis in $(0,0.1)$. The top panel and bottom panel denotes the amplitude ratio and phase shifts respectively. The color maps of amplitude ratios and phase shifts visualize $ln(R_{2})$, and $S_2$ respectively. $S_2$ and $R_2$ are defined in table \ref{tab:Shape}.}
  \label{fig:FinerEoMesh}
\end{figure}

\subsubsection{The Binary Mass Ratio}
We set up a $M_2-e_1$ grid in which $e_1$ mesh is the same as it is in the previous subsection, while $M_2$ varies on $[0.1,0.15,...1]M_{\odot}$. The binary periastron angle is set to be $\frac{1}{2}\pi$, where the amplitude ratio in both cases was shown to digress from 1 at certain eccentricities. 

\begin{figure*}[!thb]
  \centering
   \includegraphics[width=\linewidth]{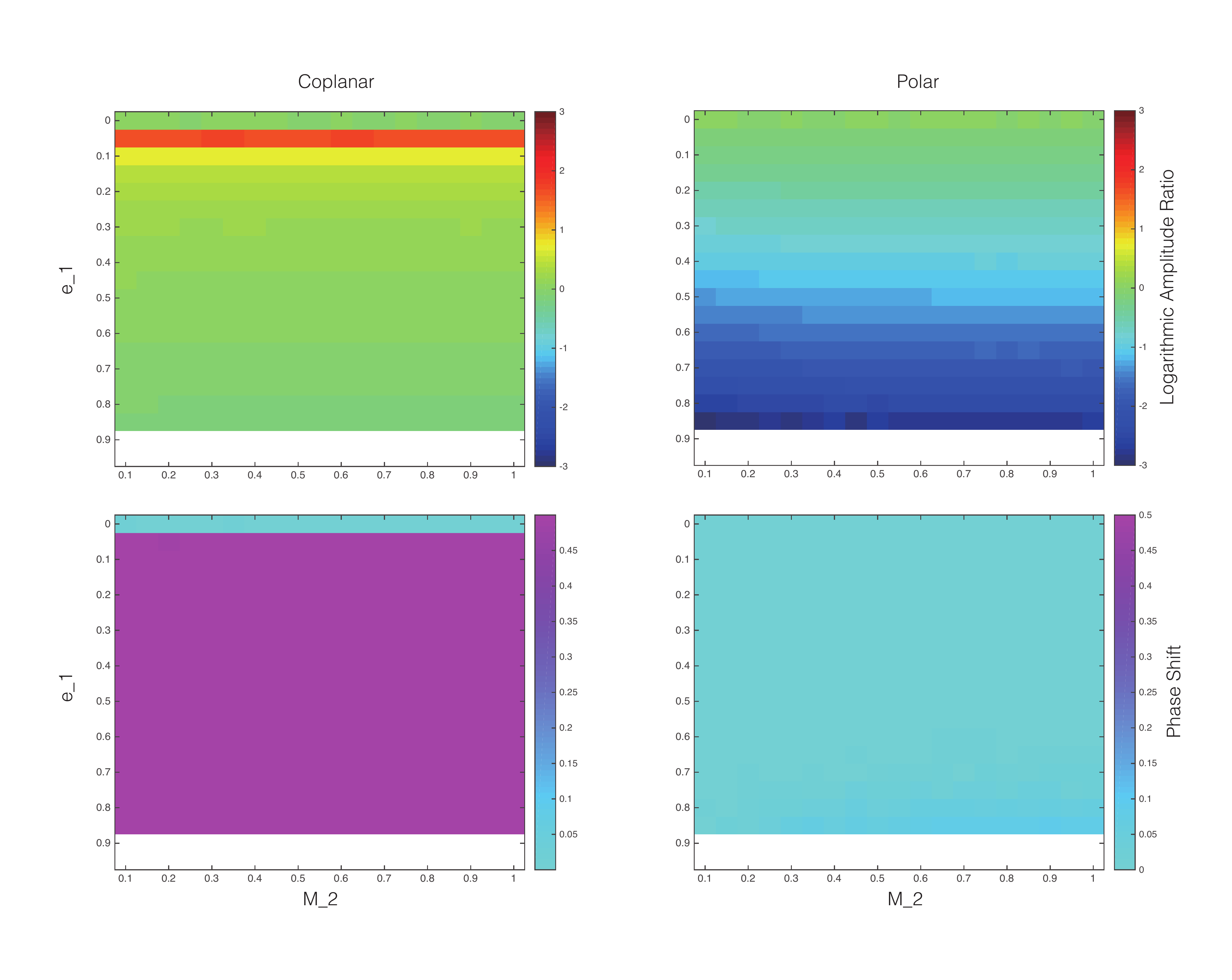}
  \caption{The maps of amplitude ratio and phase shift on the $e_1-M_2$ mesh grid. The left column is for the co-planar case, while on the right are ETVs induced by a polar planet. The top panel and bottom panel denotes the amplitude ratio and phase shifts respectively. The color maps of amplitude ratios and phase shifts visualize $ln(R_{2})$, and $S_2$ respectively. $S_2$ and $R_2$ are defined in table \ref{tab:Shape}.}
  \label{fig:eMmesh}
\end{figure*}

The results are plotted in figure \ref{fig:eMmesh}. On the x-axis, the colors in each of the figures seem to remain constant. We claim the binary mass ratio, will not give rise to any features on the maps of amplitude ratios and phase shifts. In other words, the maps should appear similar as they do in figure \ref{fig:eomesh} despite different binary mass ratios.

\subsubsection{The Outer-Inner Period Ratio}
\begin{figure*}[!thb]
  \centering
   \includegraphics[width=\linewidth]{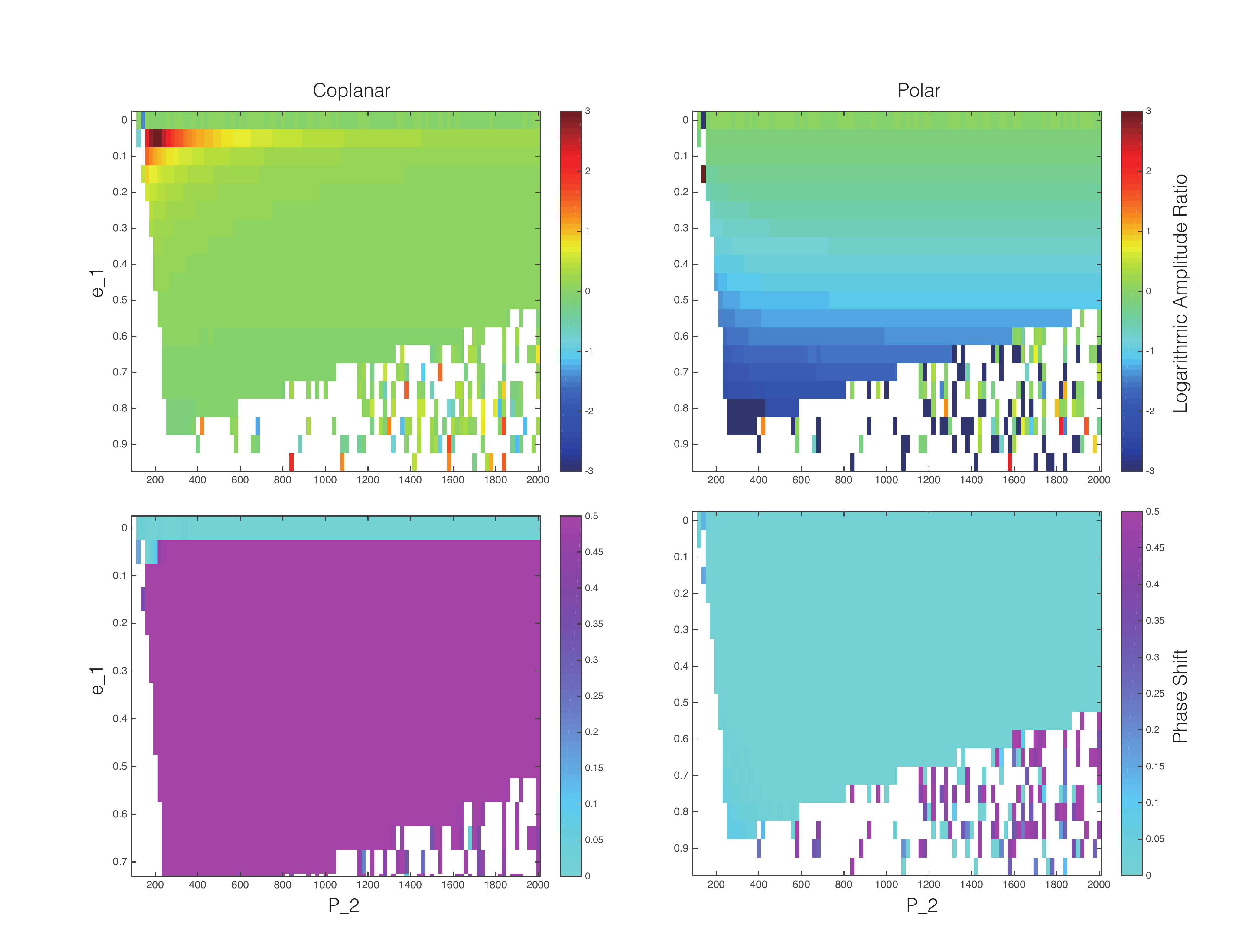}
  \caption{The maps of amplitude ratio and phase shift on the $e_1-P_2$ mesh grid. The left column is for the co-planar case, while on the right are ETVs induced by a polar planet. The top panel and bottom panel denotes the amplitude ratio and phase shifts respectively. The color maps of amplitude ratios and phase shifts visualize $ln(R_{2})$, and $S_2$ respectively. $S_2$ and $R_2$ are defined in table \ref{tab:Shape}.}
  \label{fig:ePmesh}
\end{figure*}

We set $M_2=0.5M_\odot$ and $\omega_1=0.5\pi$, where the amplitude ratio is expected to be at an extremum in either case. We let the eccentricity to be the other axis of the grid, while the grid for $P_2$ is $[100,120,140,...2000]$ (recall $P_1=27.79$~days). Note in the lowest and highest ends of this grid the planetary orbit tends to be either unstable or producing ETVs with a long stretched out tendency plus short term variations, for which the simple Fourier decomposition may not hold. Thus in the maps, we see large areas of non-results and miscellaneous color patches when the period is too short or too long. In these cases, the reliability of the results is not adequately solid. As is shown in figure \ref{fig:ePmesh}, the phase shift distinction is more obvious -- except when $e_1$ is close to 0 in the coplanar case, the phase alignment of the ETVs differs oppositely in the two categories. On the other hand, the extrema of amplitude ratios do depend on the outer period. Specifically, it seems it is only at shorter planetary periods that the abnormality in amplitudes ratios at $\omega_1=0.5\pi$ are more apparent. 

\subsubsection{The Planetary Geometric Parameters $e_2$ and $\omega_2$}
Large planetary eccentricities may have prominent impact on the ETV shapes, but we are concerned mainly with circular planetary orbits for the following reasons: (i) Eccentric CBP orbits are subject to more significant instability and are not expected to be the prevailing population. In fact, all known co-planar CBPs have their eccentricities less than 0.2. (ii) The planetary eccentricity needs to be calculated from the detailed ETVs via dynamical fitting, rather than with a priori value range from RV as for the binary, photometric observations or ETV periods as the other orbital parameters. Hence, $e_2$ and $\omega_2$ are not certain, from which we could model the ETVs produced by a polar or co-planar planet and differentiate them. Therefore, by incorporating small planetary eccentricities in the grid simulation we only aim to demonstrate whether it will change the features of co-planar and polar ETVs as shown in the previous sections.

\begin{figure*}[!thb]
  \centering
   \includegraphics[width=\linewidth]{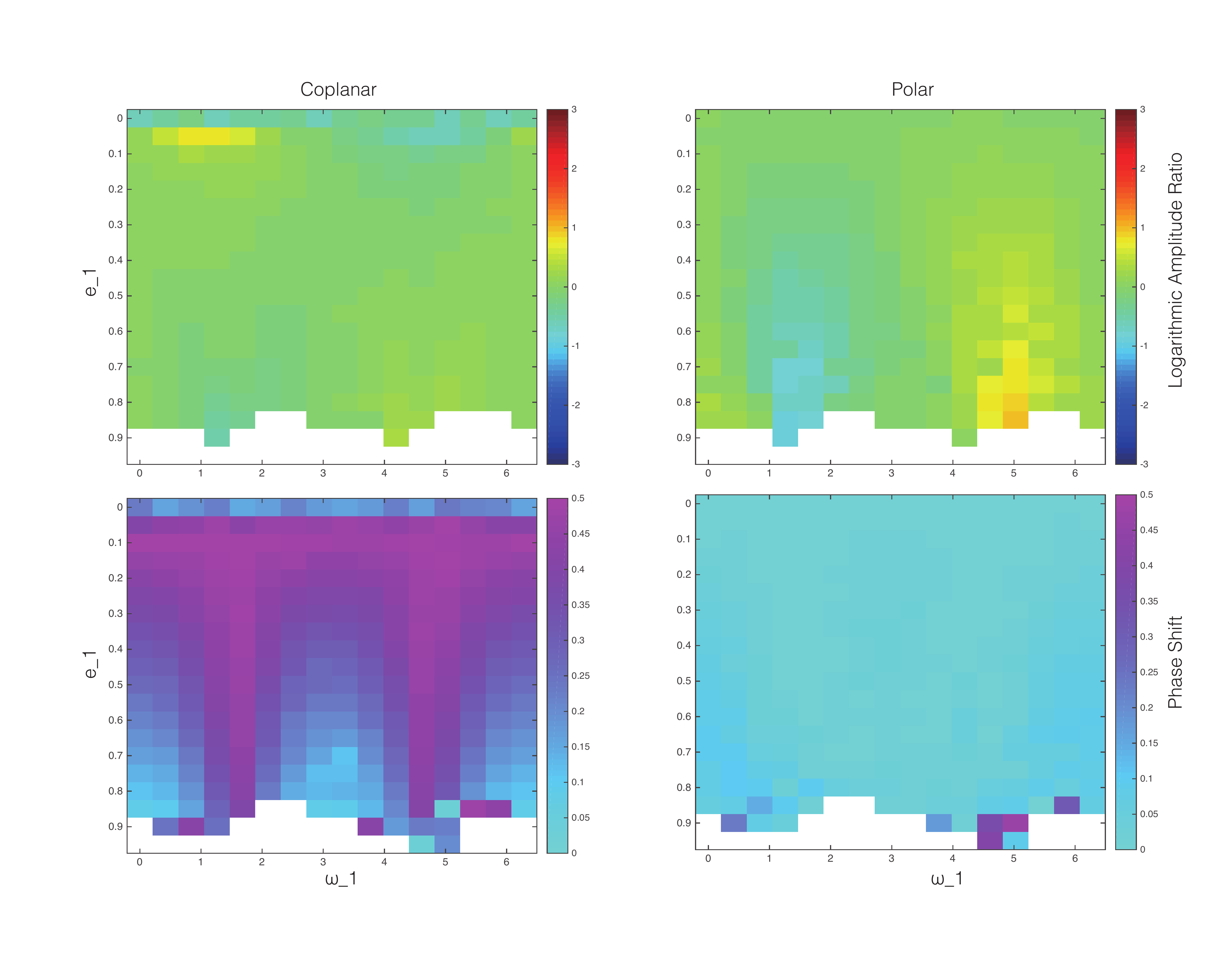}
  \caption{The maps of amplitude ratio and phase shift on the $e_1-P_1$ mesh grid, with $e_2=0.2, \omega_2=0.5\pi$. The left column is for the co-planar case, while on the right are ETVs induced by a polar planet. The top panel and bottom panel denotes the amplitude ratio and phase shifts respectively. The color maps of amplitude ratios and phase shifts visualize $ln(R_{2})$, and $S_2$ respectively. $S_2$ and $R_2$ are defined in table \ref{tab:Shape}.}
  \label{fig:e2w21}
\end{figure*}

\begin{figure*}[!thb]
  \centering
   \includegraphics[width=\linewidth]{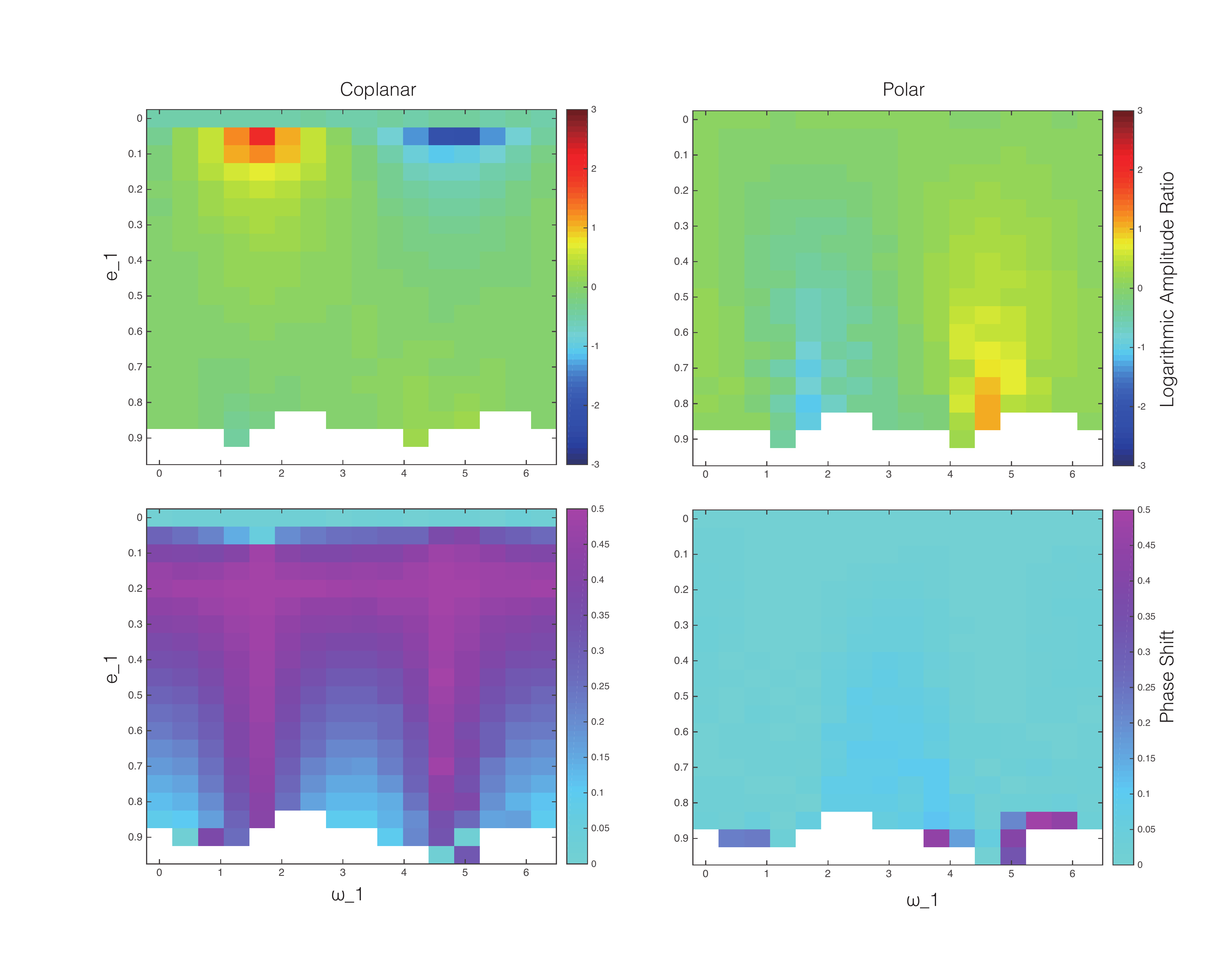}
  \caption{The maps of amplitude ratio and phase shift on the $e_1-P_1$ mesh grid, with $e_2=0.2, \omega_2=0$. The left column is for the co-planar case, while on the right are ETVs induced by a polar planet. The top panel and bottom panel denotes the amplitude ratio and phase shifts respectively. The color maps of amplitude ratios and phase shifts visualize $ln(R_{2})$, and $S_2$ respectively. $S_2$ and $R_2$ are defined in table \ref{tab:Shape}.}
  \label{fig:e2w22}
\end{figure*}

We have set $e_2=0.2$, and $\omega_2=[0, 0.5\pi]$ respectively in the simulations presented as in figures \ref{fig:e2w21} and \ref{fig:e2w22}. The remaining parameters are $M_2=0.5$ ($q=0.5$) and $P_2=300$~days ($P_2/P_1=10.8$). In these figures, the general patterns in figure \ref{fig:eomesh} remain, yet slight distortions are also visible due to the planetary eccentricity: 1), the strength of the peak and valley on the co-planar amplitude ratio map at $\omega_1=0.5\pi$ and $\omega_1=1.5\pi$ either increase or decrease, but both remain discernible. 2). The axial symmetry about $\omega_1=0.5\pi$ and $\omega_1=1.5\pi$ is warped. Despite these nuances, the overall patterns of the four maps are not destroyed by a small planetary eccentricity. 

Summarizing, in this section, our simulation on grids of the parameter space has shown that although small modification of the strengths and positions might be brought about by $P_2$ or $e_2$, the amplitude ratios and phase shifts of ETVs in a binary given its eccentricity and periastron angle are clearly different when induced by a co-planar and polar planetary companion. 

\subsection{Analytic Explanations}\label{sec:analy}
In the preceding two sections, we have demonstrated that there are various aspects of the ETV features that would be different given different geometric orbital configurations of the perturber which gives rise to it. The period discrepancy and EDV trends are due to the binary orbit precessing owing to the perturber, while the morphological features in the ETVs curves need more mathematics to be explained. In this section, we analytically and quantitatively account for the main features in the last two sections.

\subsubsection{The Precession Rate in a Hierarchical Three-Body system} \label{sec:precession}
The perturbing potential from a tertiary small object gives rise to two sorts of precessions of the binary orbit: nodal and apsidal, or in other words, the precession of the longitude of the ascending node $\Omega_1$ and the argument of periastron $\omega_1$. Both rates were calculated by \citet{Kiseleva97} and \citet{Innanen97} (later \citealt{Carruba03} found some typos in \citealt{Innanen97} and corrected them). The results are:

\begin{equation}
\label{eq:pres1}
\frac{d\omega}{dt}=\frac{G^{1/2}m_Ca_1^{3/2}}{m_A^{1/2}a_2^3}\frac{3}{4(1-e_1^2)^{1/2}}[2-2e_1^2+5(e_1^2-\sin^2i)\sin^2\omega]
\end{equation}
\begin{equation}
\label{eq:pres2}
\frac{d\Omega}{dt}=-\frac{G^{1/2}m_Ca_1^{3/2}}{m_A^{1/2}a_2^3}\frac{3\cos i}{4(1-e_1^2)^{1/2}}(1-e_1^2+5e_1^2\sin^2\omega)
\end{equation}

In both equations, $\omega$ is measured in reference to the ascending node of the planetary orbit and thus ill-defined when the planetary orbit is co-planar with the binary. However, the quantity $\varpi = \omega+\Omega$ (the longitude of periastron, also known as `\textit{pomega}') promises to be well defined and is the actual observable precessing angle. Adding equations \ref{eq:pres1} and \ref{eq:pres2}, we found that in the polar case, $\dot \Omega$ vanishes, while in the co-planar case, all terms containing ill-defined $\omega$ simply cancel, solving the definition problem. The results are:

\begin{equation}
\label{eq:presssimp1}
\frac{d(\varpi)_{cop}}{dt}=\frac{G^{1/2}m_Ca_1^{3/2}}{m_A^{1/2}a_2^3}\frac{3}{4} (1-e_1^2)^{1/2}
\end{equation}

\begin{equation}
\label{eq:presssimp2}
\frac{d(\varpi)_{pol}}{dt}=-\frac{G^{1/2}m_Ca_1^{3/2}}{m_A^{1/2}a_2^3}\frac{9}{4} (1-e_1^2)^{1/2}
\end{equation}

For an analytic expression for the period difference, we did the following computation:

We denote the precession rate $\frac{d(\varpi)}{dt}$ as $\dot{\varpi}$, and the binary radial period (time between successive minimum distances of an eccentric orbit) as $P$. Over one such period, the orbit precesses by an angle $P\dot{\varpi}$. If we assume the precession is prograde, then both the primary and the secondary eclipses would be spaced by less than $P$ by a small amount. However, the amount of time by which the primary and secondary eclipses appear in advance are different. At the time of the primary and secondary eclipses, the position vector from the secondary mass to the center of mass is 

\begin{equation}
r_{pri}=\frac{a(1-e^2)}{1+e\cos(\frac{3}{2}\pi+\omega)}=\frac{a(1-e^2)}{1-e\sin\omega}
\end{equation}

\begin{equation}
r_{sec}=\frac{a(1-e^2)}{1+e\cos(\frac{1}{2}\pi+\omega)}=\frac{a(1-e^2)}{1+e\sin\omega}
\end{equation}

Kepler's second law means the angle $P\dot{\omega}$ swept by the precessing orbit is inversely related to the distance between the two stars. This distance differs between primary and secondary eclipses, hence the difference in timing between eclipsing periods. The the ratio between the difference in triangular areas (primary vs. secondary) to the area of the whole Keplerian ellipse is the same as the ratio between the primary-secondary period difference and the orbital period ($\delta P / P$). 

Translating this logic into an equation for the primary-secondary period difference yields:
\begin{equation}
\label{eq:DeltaP}
\Delta P=\frac{\frac{1}{2}P\Delta r^2}{\pi a^2\sqrt{1-e^2}}\dot{\varpi}=\frac{2e(1-e^2)^{\frac{3}{2}}\sin\omega}{\pi(1-e^2\sin^2\omega)^2}P^2\dot{\varpi}.
\end{equation}

EDVs due to the precession of the binary orbit could be calculated in a similar manner. Suppose the impact parameter of the primary eclipse is $b$, and we define the eclipsing duration as the time between the 1st and 4th contact of the two stellar disks. The projected chord length for them to traverse would be 

\begin{equation}
\Delta L= 2\sqrt{1-b^2}(R_1+R_2).
\end{equation}

We use a rather unconventional definition for $b$ where the closest distance between the two centers is divided by the sum of the two radii, so the expressions for them in the primary and secondary eclipses are \citep{Winn10}:
\begin{equation}
b = \frac{a\cos i}{R_1+R_2}\frac{1-e^2}{1\pm e\sin\omega}
\end{equation}

where the upper sign corresponds to the primary eclipse, and lower sign is for the secondary ones.

Suppose the inclination $i$ to be close enough to $\pi/2$ so that the transverse velocity is \citep{Winn10}:
\begin{equation}
v=\frac{2\pi a}{P}\frac{1\pm e\sin\omega}{\sqrt{1-e^2}}
\end{equation}

We have the eclipsing durations given by:
\begin{equation}\label{eq:ED}
D = \frac{P(R_1+R_2)}{\pi a}\sqrt{1-\frac{a^2\cos^2i}{(R_1+R_2)^2}\frac{(1-e^2)^2}{(1\pm e\sin\omega)^2}}\frac{\sqrt{1-e^2}}{1\pm e\sin\omega}
\end{equation}

taking its derivative, and allowing for a precessing $\dot \varpi \neq 0$ binary orbit at fixed $a$ and $e$, we obtain
\begin{equation}\label{eq:EDVs}
\dot{D} = \pm\frac{P(R_1+R_2)e\cos\omega}{\pi a(1\pm e\sin\omega)^2}\frac{\sqrt{1-e^2}(1-2b^2)}{\sqrt{1-b^2}}\dot{\varpi}
\end{equation}

We now test equations~\ref{eq:presssimp1}, \ref{eq:presssimp2},  \ref{eq:DeltaP} and \ref{eq:EDVs} against simulation and observations. 

\begin{deluxetable*}{llllll}
\tablecaption{Comparison of precession rate in calculation, simulation and observation in Kepler-34 system.}
\tablehead{\colhead{\textbf{Configuration}} & \colhead{\textbf{$\Delta\varpi$ sim}}\tablenotemark{a}  & \colhead{\textbf{$\Delta\varpi$ cal}}&
\colhead{\textbf{$P_{pri}-P_{sec}$ sim}} & \colhead{\textbf{$P_{pri}-P_{sec}$ cal}} & \colhead{\textbf{$P_{pri}-P_{sec}$ obs}}} 
\startdata\hline\hline
Coplanar & 0.011$^\circ$ & 0.010$^\circ$ & 3.7s & 3.0s & 4.4s\\
Polar    & -0.202$^\circ$ & -0.181$^\circ$& -54s  & -54s  & / 
\\\hline
\enddata
\tablenotetext{a}{This denotes the total amount by which the binary orbit precesses over the simulation timespan of 4 years.}
\end{deluxetable*}\label{tab:press}

We first compare the precession rate as well as period discrepancy in table \ref{tab:press}. In both cases, it could be noted that the precession angle and period divergences obtained from simulation, calculation and observation are generally consistent, despite minor differences. The difference between the calculated precession and simulated precession might be that equations \ref{eq:pres1}, \ref{eq:pres2} only considered the quadruple moments. On the other hand, we noticed that in the simulation the period divergence seems to be inconsistent with equation \ref{eq:DeltaP} ($3.0$s vs. $3.7$s). It is not a major problem either, though, in that the precession angle in the simulation was measured by subtracting the first value from the last value in the series as shown in figure \ref{fig:CopElements}, which is subject to the short term planetary effects and not a precise value.

 However, the period discrepancies we obtained from the simulation and the observation differ, which might not be simply owing to the uncertainties in data. There are two extra sources of precession in binaries with which we could possibly solve the problem: general relativity and stellar quadrupole moments.  

The former produces  \citep{1916Einstein}:
\begin{equation}
\dot{\varpi}_{GR} = \frac{3 G^{3/2} (M_A+M_B)^{3/2}}{a^{5/2} c^2 (1-e^2)},\label{eqn:gr}
\end{equation}
which is 0.007 degrees every 4 years (the \emph{Kepler} data span) for the binary parameters as given in Table~\ref{tab:K34params}, which is comparable with the value obtained from Newtonion simulations and calculations (see table \ref{tab:press}). Considering the extra precession induced by GR would result in a calculated period discrepancy of $\sim$ 4.8s, closer to the observation. Adding it to the simulation would yield 5.5s, rather beyond the observational value.  

The latter source of precession is from the stellar quadruple moment, whcih depends on tides and rotation. The strength of these effects depend on the internal central concentration of the stars as summarized by the apsidal motion constant $k_2$ \citep{1939Sterne}. Values from \cite{1995Claret} for stars like A and B in Kepler-34 have values in the range $k_2=0.013-0.022$. The precession rate may be written: 
\begin{eqnarray}
    \dot{\varpi}_T &=& 30 \pi \frac{G^{1/2} (M_A+M_B)^{1/2} }{ a^{13/2} } \frac{1 + (3/2)e^2+(1/8)e^4}{(1-e^2)^5} \nonumber \\
    &&\times ( \frac{M_B}{M_A} k_{2A} R_A^5 + \frac{M_A}{M_B} k_{2B} R_B^5 ), \label{eqn:tides}
\end{eqnarray}
and we find a rate of $0.0020-0.0034$ degrees per 4 years. This is smaller than the GR contribution, and it was hence neglected by \cite{Welsh12} in analyzing the system, but it is not entirely negligible. 

The individual stellar spins of angular velocity $S$ also produces a quadrupole potential, causing a component of in-plane precession. The value is: 
\begin{eqnarray}
    \dot{\varpi}_R &=&  \frac{[(M_A+M_B)/G]^{1/2} }{ 4 a^{7/2} (1-e^2)^2 } \nonumber \\
    &&\times ( \frac{k_{2A}}{M_A} S_A^2 (5\cos^2\psi_A-1) R_A^5  \nonumber \\
    && +\frac{k_{2B}}{M_B} S_B^2 (5\cos^2\psi_B-1) R_B^5 ), \label{eqn:spinprec}
\end{eqnarray}
where we include the effect of obliquity, in which the stellar spin axis makes an angle $\psi$ to the orbital angular momentum. Fast-spinning, high-obliquity stars can cause retrograde precession, as demonstrated by \cite{2009Albrecht} for a system in which a third body had previously been invoked to explain the anomalous precession \citep{2006Hsuan}. Nevertheless, if the stars are spinning near the pseudo-synchronous rate of \citep{1981Hut}:
\begin{equation}
    S_{ps} = \frac{2 \pi}{P} \frac{1+(15/2)e^2+(45/8)e^4+(5/16)e^6}{(1+3e^2+(3/8)e^4)(1-e^2)^{3/2}},
\end{equation}
then the magnitude of the spin contribution is smaller than the tidal contribution. For Kepler-34, this value is $S_{ps} = 0.682$day$^{-1}$, and quasi-periodic variability suggests the actual rotation angular velocities of the two components are between $0.35$ and $0.42$day$^{-1}$ \citep{Welsh12}, i.e. tides are too weak to pseudo-synchronize the components and they spin down via magnetic breaking.  We suppose, therefore, that $\Omega_{ps}$ is a good upper bound on $\Omega_A$ and $\Omega_B$, and $0.022$ is an upper bound on $k_2$, leading to evaluation of equation~\ref{eqn:spinprec} of $\lesssim1.3\times10^{-5}$~deg over 4 years for Kepler-34. So the rotation contribution is negligible compared with the GR contribution.

To sum up, taking the GR into account helps ease the difference of period divergence in simulation without GR, and observation, but not completely. The inclusion of the tidal effect, which was not done when deriving the model in \citet{Welsh12}, may hint that the model suffers systematic errors in planetary parameters such as the eccentricity, which eventually leads to the difference between our simulation and the observation. But all these subtle inconsistency would not qualitatively change our conclusion that by computing the period divergence and comparing it with the observed values, a polar and a coplanar planet could be effectively distinguished.

Finally, we calculated the total amount of EDV over the simulation span to test equation~\ref{eq:EDVs}. In the coplanar case, it yields $\sim-0.4s$ and $\sim2.8s$ for the primary and secondary eclipses, while in the polar case, we obtained $\sim-6s$ and $\sim52s$ respectively, i.e., in good consistency with the simulation shown in figures \ref{fig:CopEDV} and \ref{fig:PolEDV} where we plotted black dash lines to represent the analytic predictions.

\subsubsection{Shape of the ETV Curves}
In this section our aim is to explain the patterns present in figure \ref{fig:eomesh}, i.e, to at least qualitatively explain why in most cases the phases of the primary and secondary ETVs match in the polar case, while in the co-planar case they are more likely to be opposite one another.  

We turn to the analytic expressions of the ETVs provided with all the orbital parameters in the CBP system given by \citet{Borkovits1} using perturbation theory. It turned out fruitless to study the ETV shapes with the entire expression, which is too long for any physical content to stand out. To divide and conquer, we resorted to \citet{Borkovitz2}, where the authors divided the ETVs into several terms with different mechanisms and time scales.

\begin{figure}[!thb]
  \centering
   \includegraphics[width=\linewidth]{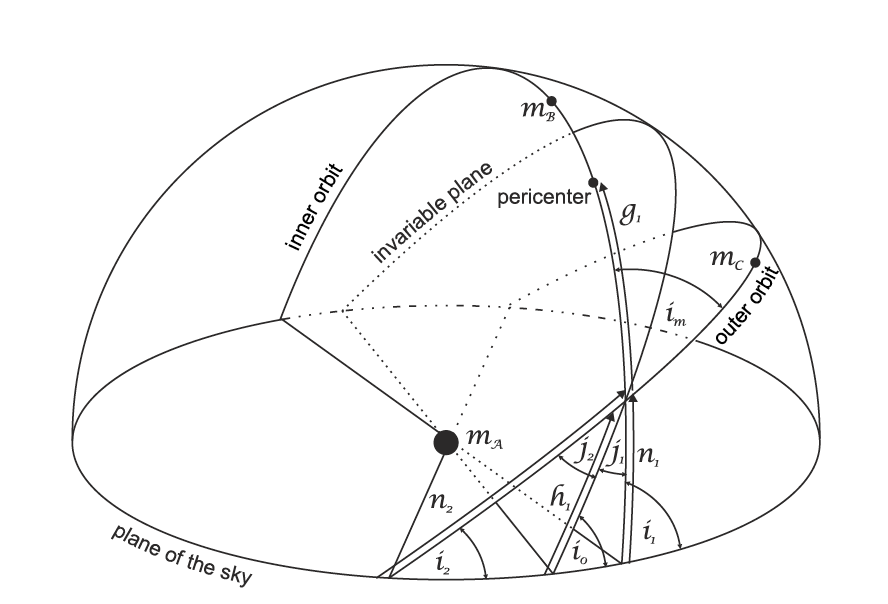}
  \caption{Orbital configuration of a hierarchical triple system, from \citet{Borkovitz2}.}
  \label{fig:config}
\end{figure}

To make the denotations of \citet{Borkovitz2} clearer, in table \ref{tab:marks} and figure \ref{fig:config} we copied the table that contains the meanings of all the parameters they used, and a diagram visualizing the system and the parameters. We have kept our notations consistent with that table.

\begin{deluxetable}{llll}
\tablecaption{Orbital parameters in a hierarchical triple system, from \citet{Borkovitz2}}
\tablehead{\colhead{Parameter} & \colhead{Symbol} & \colhead{Explanation}}
\startdata
Mass\\
Binary Members & $m_{A,B}$ & \\
Total mass of the binary & $m_{AB}$ & $m_{A}+m_{B}$\\
Tertiary mass & $m_{C}$ &\\
Total mass & $m_{ABC}$ & $m_A+m_B+m_C$ \\
\hline
Period & $P_{1,2}$ &\\
\hline
Semi-major axis & $a_{1,2}$ & \\
\hline 
Eccentricity &$e_{1,2}$&\\
\hline
Mean anomaly & $l_{1,2}$ &\\
\hline
True Anomaly & $v_{1,2}$ &\\
\hline
True Longitude & $u_{1,2}$ &\\
\hline
Argument of Periastron & $\omega_{1,2}$ &\\
\hline
Inclination \\
Observable & $i_{1,2}$ &\\
mutual & $i_{m}$ &\\
&I&$cos i_{m}$\\
\hline
Ascending Node\\
obsevational &$\Omega_{1,2}$ &\\
sky-dyn. nodes angle & $n_{1,2}$ &\\
&$\alpha$ & $n_{2}-n_{1}$ \\
&$\beta$ & $n_{2}+n_{1}$ \\
\hline
\enddata
\end{deluxetable}\label{tab:marks}

The ETVs could be divided into three terms ranging from the weakest to the strongest:

\textbf{1)} The \textit{Light Travel Time Effect (LTTE)}. This term has the period of the planet, but its scale is far weaker than the other two terms. In the case of Kepler-34, it would result in ETVs of the amplitude $0.055$s, which can be safely neglected as it is not only much smaller than the uncertainties we could acquire from Kepler timing measurements, but also much smaller than the dynamical effect, thus not affecting their relative shapes.

\textbf{2)} Dynamics on $P_2$ time scale. It is the main oscillating term observed.

\textbf{3)} The apse-node time effects, which reflects the precession of the triple system as a whole. It is of much longer time scale compared to the simulation time-span, and thus, although it is the strongest effect, we can encapsulate it into a background linear effect, as in equation \ref{eq:DeltaP}.

When focused on the planetary dynamics effect (2), we are still faced with an infinite number of terms in the expansion of $\frac{P_2}{P_1}$. But mainly three terms come into effect: \textbf{2a)} the quadruple term; \textbf{2b)} the octuple term; and \textbf{2c)} what the authors of \citet{Borkovitz2} named as \textit{$P_2$ time-scale residuals of the $P_1$ time-scale dynamical effects}. The \textbf{2b} term is smaller than \textbf{2a)} by a factor of $\frac{1-q}{1+q}$, while the \textbf{2c} term is smaller by a factor of $\frac{P_1}{P_2}$. We would focus on the simplest case in which the binary mass ratio is close enough to 1 so that $\frac{1-q}{1+q} \ll 1$, and the period ratio is sufficiently small, when the second and third term vanish. Both assumptions are solid in the Kepler-34 system, but not necessarily as good in our simulation settings for figure \ref{fig:eomesh} (where $q=0.5$, $P_1/P_2=0.0926$). 

The \textbf{2b} term could be simplified when the geometrical parameters are given. We derived the corresponding simplified forms of it in the co-planar and static polar configurations. Before presenting them, we first define a few constants and coefficients.
\footnote{Expansions of coefficients $f_1, K_1, K_{11}$ and $K_{12}$ to even higher order could be found in the appendix of \citet{Borkovitz2}.} 
\begin{equation}
\begin{aligned}
A_{L1}=&\frac{15}{8}\frac{m_C}{m_{ABC}}\frac{P_{1}}{P_{2}}(1-e^{2}_{2})^{-3/2}\\
M=&v_{2}-l_{2}+e_{2}\sin v_{2}\\
S(2u_{2})=&\sin 2u_{2} +e_{2}[\sin(u_{2}+\omega_{2})+\frac{1}{3}\sin(3u_{2}-\omega_{2})]\\
C(2u_{2})=&\cos 2u_{2} +e_{2}[\cos(u_{2}+\omega_{2})+\frac{1}{3}\cos(3u_{2}-\omega_{2})]\\
 f_1=&1+\frac{25}{8}e^{2}_{1}+\frac{15}{8}e^{4}_{1}+\frac{95}{64} e^{6}_{1} + O(e_{1}^{8}) \\
 K_1 =& \mp e_{1}\sin\omega_1+\frac{3}{4}e_{1}^{2}\cos2\omega_{1} + O(e_{1}^3)\\
 K_{11}=&\frac{3}{4}e^2_1 \pm e_1\sin \omega_1+\frac{51}{40}e^2_1\cos \omega_1 +O(e_1^3) \\
 K_{12}=&\mp e_1 \cos \omega_1 +\frac{51}{40} e^2_1 \sin 2 \omega_1+O(e_1^3) 
\end{aligned}
\end{equation}

in which the upper signs are for the primary eclipses, while the lower are for the secondary eclipses.

Then we present the simplified $O-C$ expressions in the two cases.

\textbf{Coplanar}\par
 In this case the expression reduces to the form:
\begin{equation}
 \begin{aligned}
 O-C = &\frac{P_1}{2\pi}A_{L1}\sqrt{1-e^2_1}([\frac{8}{15}+\frac{4}{5}K_1]M+\\&
 K_{11}S(2u_{2}-2\alpha)-K_{12}C(2u_{2}-2\alpha)))
  \end{aligned}
 \end{equation}
 which could be further reduced in the first order of $e_1$:
  \begin{equation}\label{eq:13}
 \begin{aligned}
 O-C &\approx\frac{P_1}{2\pi}A_{L1}\sqrt{1-e^2_1}([\frac{8}{15}+\frac{4}{5}K_1]M\\&\pm2e_1C(2u_2-\omega_1)+O(e_1^2))
 \end{aligned}
 \end{equation}
\textbf{Polar}\par
We expand the results into the following form:
\begin{equation}
\begin{aligned}
O-C = &\frac{P_1}{2\pi}A_{L1}\sqrt{1-e^2_1}(B_{0}M+\\
      &\sum^{3}_{j=1}(A_i \cos ju_2+B_i \sin ju_2))
\end{aligned}
\end{equation}

where we have derived the coefficients before each of the 'Fourier' harmonics (its argument $u_2$ is non-linear in time unless the planetary eccentricity vanishes, thus would not be consistent with $F_i$ presented in table \ref{tab:Shape}):
\begin{equation}
\begin{aligned}
B_{0}=&-\frac{4}{15}f_1-\frac{2}{5}+2K_1\cos2\omega_1; \\
A_1=&(3K_{11}\cos 2\omega_1-\frac{2}{3}f_1-K_1)e_2\cos\omega_2+\\
&2(K_{11}\cos\omega_1-K_{12}\sin\omega_1)e_2\sin\omega_2;\\
B_1=&-(K_{11}\cos2\omega_1+\frac{2}{15}f_1+\frac{1}{5}K_1)e_2\sin\omega_2\\
&+2(K_{12}\sin\omega_1-K_{11}\cos\omega_1)e_2\cos\omega;\\
A_2=&-\frac{2}{5}f_1-\frac{3}{5}K_1-K_{12}\sin2\omega_1; \\
B_2=&0;\\
A_3=&\frac{1}{3}e_2[(K_{11}\cos2\omega_1-\frac{2}{5}f_1-\frac{3}{5}K_1)\cos\omega_2\\
&+2(K_{11}\cos3\omega_1-K_{12}\sin3\omega_1)\sin\omega_2];\\
B_3=&\frac{1}{3}e_2[(\frac{2}{5}f_1+\frac{3}{5}K_1-K_{11}\cos2\omega_1)\sin\omega_2\\
&+2(K_{11}\cos3\omega_1-K_{12}\sin3\omega_1)\cos\omega_2];
\end{aligned}
\end{equation}

First and third order harmonics as well as $M$, scale with the planetary eccentricity, thus vanish when the planetary orbit is assumed to be circular, as is in the simulation presented by figure~\ref{fig:eomesh}. Taking this as an extra assumption, in both coplanar and polar geometries the ETV curves would be purely sinusoidal with frequency double planetary orbital frequency. 

In the co-planar case, the only term remaining is $\pm e_1\cos(2u_2-\omega_1)$. The primary and secondary ETVs only differ in the sign of the coefficient -- which indicates that the primary and secondary ETVs are of the same amplitude, but phases shifted by $\pi$, as was seen in figure \ref{fig:eomesh}. It is consistent with what is seen in Kepler-34 (whose planetary eccentricity is close enough to zero that this picture holds qualitatively) and consistent with most parts of figure \ref{fig:eomesh}, except the `peak' and `valley' in low inner eccentricity region, which will be revisited later.

\begin{figure}[!thb]
  \centering
   \includegraphics[width=\linewidth]{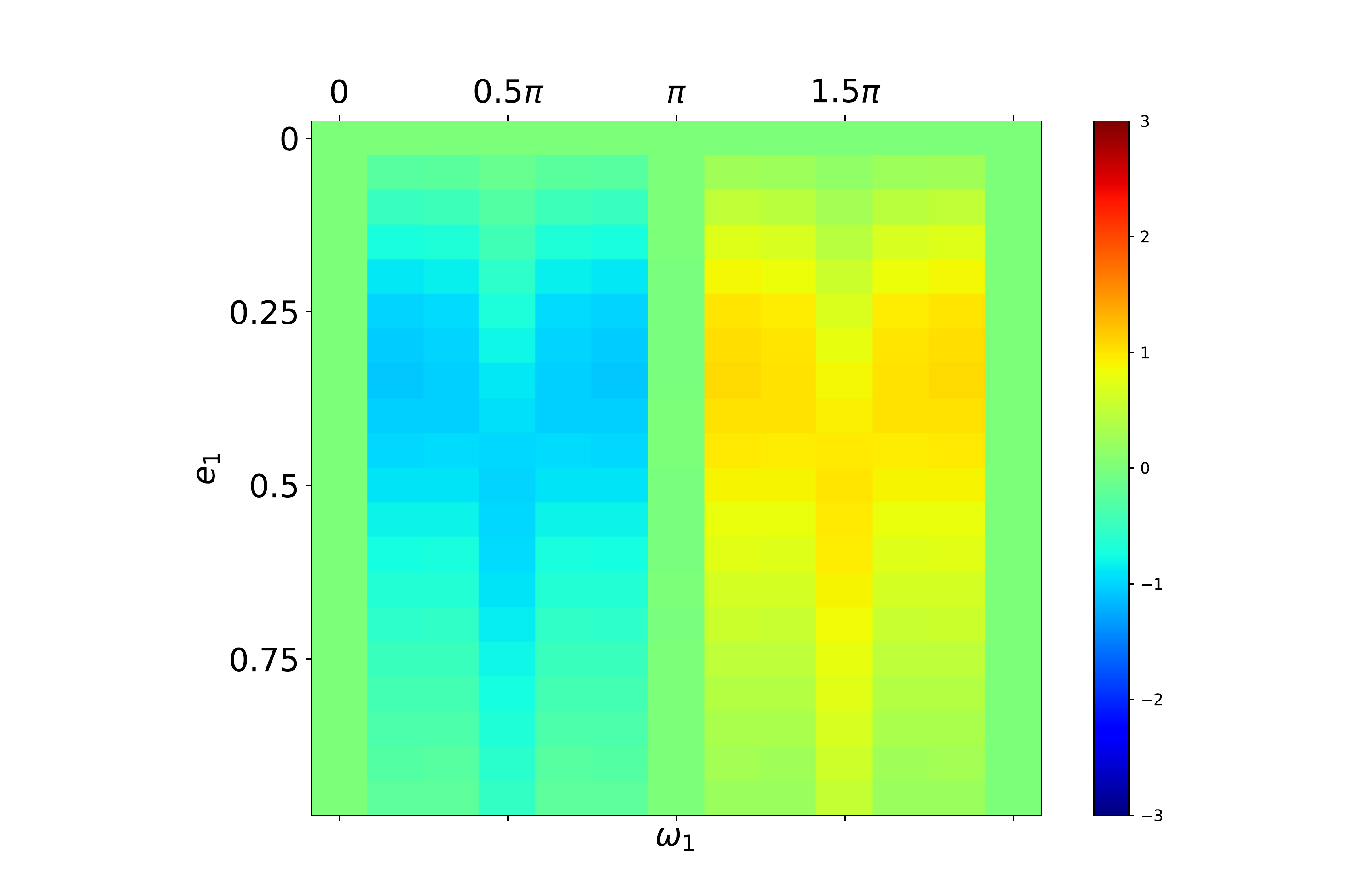}
  \caption{The amplitude ratio of ETVs induced by a polar companion according to the analytic expression on $e_1-\omega_1$ mesh grid. The colorbar represents $ln(R_2)$, in which $R_2$ is defined as in table \ref{tab:Shape}. Note to draw this figure we used the expressions of $K_1,K_{11}$ and $K_{12}$ to the 7th-order of $e_1$, which could be found in \citet{Borkovitz2}. }
  \label{fig:ana_ratiop}
\end{figure}

In the polar case we need to evaluate $A_2$ as a function of $e_1$ and $\omega_1$ for the primary and secondary ETVs, and compute the ratio of them to compare with figure \ref{fig:eomesh} on the same mesh grid. We obtained figure \ref{fig:ana_ratiop}. In order to be as accurate as possible, we adopted the expression for $f_1,K_1,K_{11}$ and $K_{12}$ to the seventh order of $e_1$ from \citet{Borkovitz2}. Comparison between figure \ref{fig:ana_ratiop} and figure \ref{fig:eomesh} is qualitatively fairly consistent when $e_1$ is not too large, i.e, in figure \ref{fig:ana_ratiop} the amplitude ratio of ETVs diverges from unity with increasing inner eccentricity. While when $e_1$ gets larger than 0.5, in figure \ref{fig:ana_ratiop} the amplitude ratio does not continue to increase as was shown in figure \ref{fig:eomesh}. In terms of the phase shift, the analytic prediction is that the phases would always be aligned in the polar case, also qualitatively consistent with the lower eccentricity region of figure \ref{fig:eomesh}.

To sum up, the \textbf{2b} term in the ETV expressions alone could qualitatively account for most of the patterns present in figure \ref{fig:eomesh} in both configurations. However, it does not sufficiently explain (i) the tendency of amplitude ratio to continue diverging from unity when $e_1$ increases beyond $\sim 0.3$ in the polar case, or (ii) the color patch in the amplitude ratio map of co-planar ETVs at lower eccentricities. The first puzzle we will not attempt to solve with more terms in $e_1$, which would converge slowly. The second puzzle, on the contrary, is easily solved by taking previously-ignored terms into account, as follows. 

The \textbf{2b} term vanishes with the inner eccentricity in the co-planar case. To explain the abnormalities when $e_1$ is small in the coplanar case, terms previously neglected needs to be reinstated. In the co-planar case the second harmonics in \textbf{2b} term vanish with $e_1$, while according to equation 20a in \citet{Borkovitz2}, \textbf{2c} has the only non-vanishing second 'harmonic' component, proportional to $-\frac{11}{30}\sin(2u_2)$. It would be of comparable magnitude with \textbf{2b} term in the low inner eccentricity region, and the relative sizes of them determine the amplitude ratio and phase shift of ETVs. Adding them together, we calculate the amplitude ratio of the primary and secondary ETVs approximated to the first order of $e_1$ as

\begin{equation}\label{eq:ratioc}
R_{pri/sec} = \frac{\sqrt{e_1^2+\frac{121 P_1^2}{900 P_2^2}+\frac{11P_1}{15P_2}e_1\sin\omega_1}}{\sqrt{e_1^2+\frac{121 P_1^2}{900 P_2^2}-\frac{11P_1}{15P_2}e_1\sin\omega_1}}
\end{equation}

\begin{figure}[!thb]
  \centering
   \includegraphics[width=\linewidth]{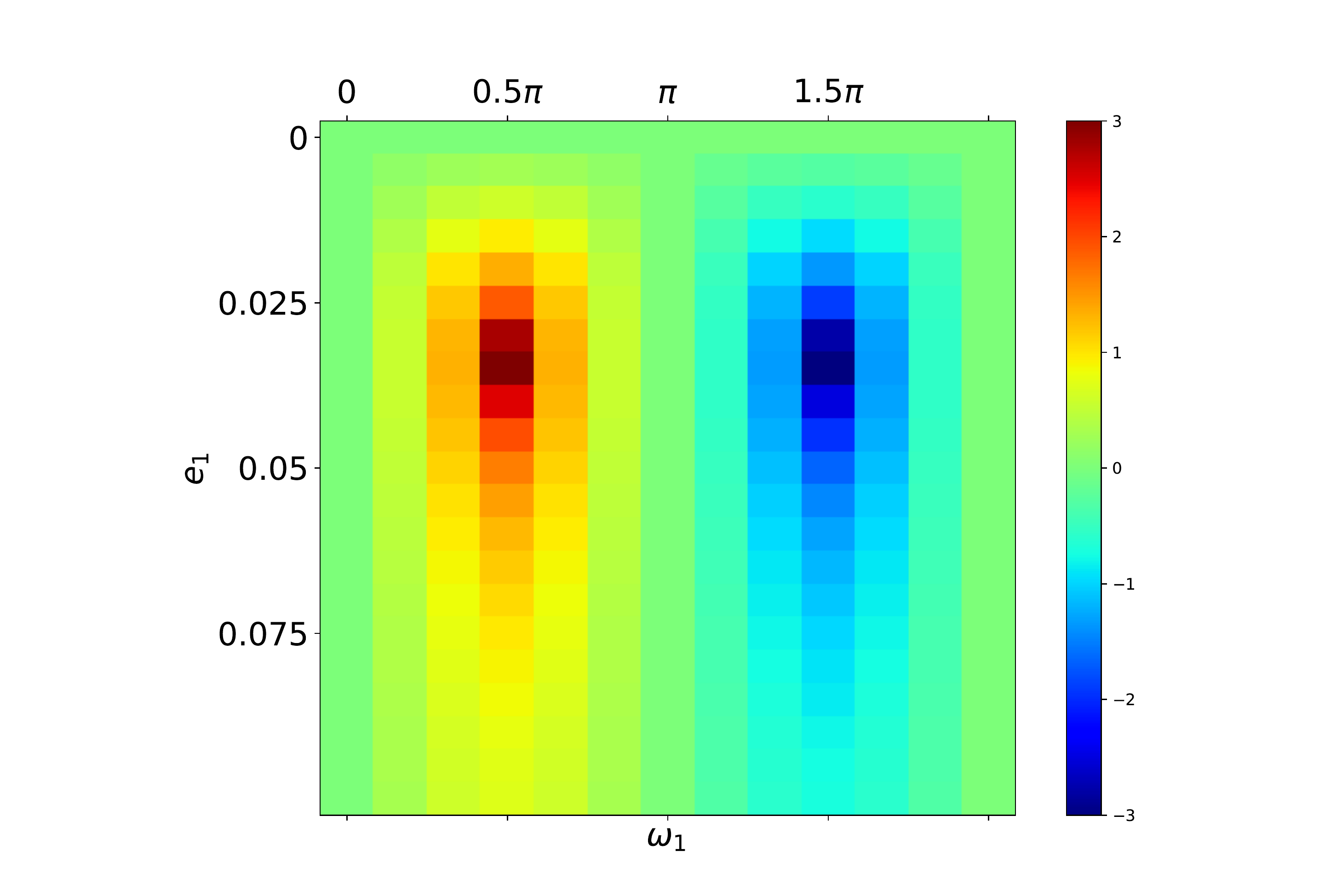}
  \caption{The amplitude ratio of ETVs induced by a co-planar companion according to equation~\ref{eq:ratioc} in the $e_1-\omega_1$ plane, in the small $e_1$ region. The colorbar represents $ln(R_2)$, in which $R_2$ is defined as in table \ref{tab:Shape}. }
  \label{fig:ana_ratioc}
\end{figure}

Figure \ref{fig:ana_ratioc} is this expression as a function of $e_1$ and $\omega_1$ on the same grid as in the left column of figure \ref{fig:FinerEoMesh}. They agree with each other perfectly. Equation \ref{eq:ratioc} indicates that at $\omega_1=0.5\pi$ or $\omega_1=1.5\pi$, an extremum of the amplitude ratio would occur at $e_1=\frac{11P_1}{30P_2}$, which when $P_1=27.79$ and $P_2=300$ equals to $\sim0.03$, agreeing with figure \ref{fig:FinerEoMesh}. 

\section{Discussions}\label{sec:diss}

\subsection{Application}
The ultimate aim of our work is to enable researchers to identify the orbital orientation of a planet orbiting a binary, using only the ETV phenomena. In this subsection, we summarize how it can be done, as well as discuss some restrictions on the applicability of our suggested methods.

The three major distinctions are:

\begin{enumerate}
    \item \textbf{The period difference}. The primary eclipses and secondary eclipses have different periods. We refer the reader to equations \ref{eq:pres1}, \ref{eq:pres2}, \ref{eq:presssimp1}, \ref{eq:presssimp2} and \ref{eq:DeltaP}. By analyzing the eclipsing timings themselves or by radial velocities, $e_1$ and $\omega_1$ can be obtained. A simple criterion goes as: \textbf{If $\sin\omega_1$ is positive, for a polar CBP, the secondary eclipsing period of the binary is longer than that of the primary, while for a coplanar one, we would expect a longer secondary period. For a negative $\sin\omega_1$, vice versa.} This criterion is, of course, just the contribution to apsidal motion from the planet; additional precession terms, equations~\ref{eqn:gr}, \ref{eqn:tides}, and \ref{eqn:spinprec}, also need to be considered. 
    \item \textbf{Eclipse Duration Variations (EDVs)}. Generally, EDVs are harder to measure than ETVs or period differences. But if available, according to equation \ref{eq:EDVs}, they provide another simple criterion. \textbf{If $\cos\omega_1(1-2b^2$) is positive, in the coplanar case, the primary eclipsing duration decreases while the secondary ones increase, in the polar case, primary eclipses get longer while secondary ones get shorter. Vice versa when $\cos\omega_1*(1-2b^2)$ is negative.} In addition, the rate of EDVs is indicative of different planetary masses in the two models, and one of them could be ruled out with the constraints from the ETV amplitudes.
    \item \textbf{Comparing primary and secondary ETV}. We have shown in section \ref{sec:grid} that the amplitude ratio and phase shift between primary and secondary ETV mostly depend on $e_1$ and $\omega_1$ qualitatively, while other parameters of the planet's orbit, of which we do not have any prior knowledge, could distort, reduce, or intensify the patterns on the $e_1-\omega_1$ map. Therefore, less trust should be invested in when using the curve morphology as a configuration indicator. \textbf{Generally, primary and secondary ETVs at twice the frequency of the planet's orbital frequency tend to be aligned in phase for polar planets, while they tend to be anti-aligned in the co-planar case.} On the other hand, given a large binary eccentricity $e_1$, the amplitude ratio being far from 1 probably indicates a polar planet. Given a small binary eccentricity $e_1$, if the amplitude of the primary and secondary ETVs turns out far from 1 it may indicate a coplanar planet. Reliability of these criteria also depends on $\omega_1$, since the patterns in figure \ref{fig:eomesh} are most emphasized at $\omega_1=\pi/2$ or $\omega_1=3\pi/2$. More reliably, one should locate the binary on $e_1-\omega_1$ plane, and compare the color block in figure \ref{fig:eomesh} with the observed values.
\end{enumerate}

\subsubsection{To what extent can the configuration be hinted from the ETVs?}
We have focused on two groups of the planetary orbit among a much larger parameter space. According to \citet{Farago10} and \citet{Doolin11}, the stable misaligned planetary orbits are often topologically identified with the static polar orbit; for example, the planetary angular momentum could be misaligned by a small angle with respect to the binary semi-major axis, or the mutual inclination differs slightly from $\pi/2$. Meanwhile, although for highly inclined binary, the 'topologically polar' orbits could have considerably low inclinations, the formation scenario in which the binary eccentricity drives the initially inclined disk all the way into the polar state would prefer the end state of this evolution -- the exact polar states. In these cases, a static polar orbit would be a good approximation for them, and we expect the ETVs to behave analogously, just like when talking about `co-planar' we also include orbits that are inclined by a few degrees. The largest mutual inclination among well-constrained CBPs is $4.1^{\circ}$ \citep{Kostov14}, and this value is small enough for the ETVs to be almost identical between the actual case and its coplanar counterpart. 

In this sense, the distinctions we have found could help one tell the geometric category of the planet only if it orbits close enough to the standard model of these two categories. Moderately misaligned CBPs, if they exist, would complicate the result. A planet misaligned by only $15^{\circ}$ might give rise to ETVs consistent with a co-planar planet. Therefore, what we find could be used to distinguish a polar planet from the co-planar ones, but may not be able to select out all misaligned CBPs.  

We would like to remark some requirements of the systems for our criteria to be applied. First, the stronger two of the three criteria, the ones relying on ETVs, are concerned with joint features of the primary and secondary eclipses. Thus, binary systems with two stars of similar temperature, where primary and secondary eclipses could be discerned and measured with similar accuracy, are the better subjects to apply the criteria to. Second, the planetary period should not be too long so that the relative phase of the primary and secondary ETVs could emerge during the observation mission time span. Third, the binary eccentricity $e_1$ needs some subtle consideration. When $e_1$ is small, the abnormality in figure \ref{fig:eomesh} appears but they would be hard to measure as ETVs would be of much smaller amplitude. In addition, in these cases, the double frequency term would be less representative of the whole ETV curves, since first-order terms would be of comparable strengths. On the other hand, larger inner eccentricity means not only larger ETVs, but also less pronounced distinctions in the ETV shapes, as shown in figure~\ref{fig:eomesh}. To sum up, moderate inner eccentricities are preferred when applying the criteria.

\subsection{Perspective}
Dynamical ETVs is a method that promises to be useful for surveying the statistics of orbital inclinations of CBPs, because the amplitude of is not strongly dependent on orbital orientation. Other methods are biased towards coplanar planets.  Light-time effect ETVs or radial velocities linearly depend on $\sin i$, and hence for eclipsing binaries they are less sensitive to polar planets than they are to coplanar planets. The transit method is biased towards very coplanar systems, because those systems show a comprehensible series of planet transits, rather than sparse signals in the case of highly inclined planets, which are hard to notice and harder to invert for orbital parameters.

There are about 3000 eclipsing binaries in the \textit{Kepler} field according to the \textit{Villanova Kepler Eclipsing Binary Catalog} \citep{Batalha11}, many of which have precise enough ETV measurements but are not completely analyzed. Detecting non-transiting CBPs with the help of ETVs in the \textit{Kepler} database is a task far from accomplished, in the remaining process of which our study could help detect non-coplanar CBPs. Besides, TESS \citep{Ricker14} has and will be discovering a considerable number of eclipsing binaries. Although the observation duration of TESS towards a certain object might be much shorter, we could still expect to know some eclipsing binaries with sufficiently precise measurements in eclipsing timings and durations. Our method will be
an effective way to explore these datasets, broadening the catalog of misaligned CBPs, as long as sufficient measurements of eclipsing timings both primary and secondary are available. Further CBP statistics studies using \textit{PLATO} \citep{PLATO} will be also availed by our study. Detections or non-detections of more mis-aligned CBPs will surely raise more theoretical questions on the formation mechanisms and more aspects of them.

\acknowledgments
We thank our colleagues, particularly T. Borkovits and D. Martin, for helpful conversations. ZB acknowledges the support from Xuening Bai and Institute for Advanced Study at Tsinghua University.  DF's work has been supported by NASA under Grant No. NNX14AB87G (Kepler Participating Scientist Program) and NASA-NNX17AB93G (Exoplanet Research Program).

\vspace{5mm}
\facilities{Kepler}

\software{\texttt{Matlab, python3, numpy, scipy, rebound, emcee, IDL, EXOFAST}}

\appendix

 The planet in Kepler-34 was easily detected, and well constrained, with just the first part of the \emph{Kepler} dataset, hence the full set of \emph{Kepler} data have not been analyzed in the published literature. Therefore we decided to derive the remaining eclipse times. We downloaded the lightcurve data from MAST\footnote{\url{https://archive.stsci.edu/}; we used release 20-23, depending on quarter.} and used SAP flux, short cadence where available (quarters 13-17) and long cadence elsewhere (quarters 1-12), for KIC 8572936.  We used the \texttt{occultnl} model from \cite{Mandel02}, as implemented by \cite{2013Eastman}, to model the eclipses. The relative positions of each star during eclipse was approximated as rectilinear motion over one another with a given impact parameter and timescale --- one pair for primary and one pair for secondary eclipse. Each eclipse was diluted by the flux of the other star --- flux ratio was a fitting parameter --- as well as a seasonal contamination values given on the MAST website. Stellar radii relative to each other, quadratic limb darkening coefficients for each star, and mid-time of each eclipse, were also fitting paramters.  After dividing the data by the model, a best-fit curve based on a cubic polynomial fit to $\pm0.75$~days around each eclipse was subtracted, to handle instrumental and slow astrophysical drifts.  These fits were obtained with mpfit in IDL \citep{2009Markwardt}, and the eclipse mid-times and their formal error bars are given in tables~\ref{tab:PETVs} and \ref{tab:SETVs}.

\begin{deluxetable*}{lll}
\centering
\caption{Primary eclipse timings of the Kepler-34 binary.}\label{tab:PETVs}
\tablehead{
\colhead{Index} & \colhead{Timing (BJD-2454900)} & \colhead{Uncertainty (day)}}
\startdata
      0.0  &      79.7239087 &  0.0000154\\
      1.0  &     107.5196598  & 0.0000180\\
      2.0  &     135.3153336 &  0.0000187\\
      3.0  &     163.1111874 &  0.0000190\\
      4.0  &     190.9069779 &  0.0000201\\
      5.0  &     218.7028811 &  0.0000159\\
      6.0  &     246.4987712 &  0.0000161\\
      7.0  &     274.2944583 &  0.0000163\\
      8.0  &     302.0901904 &  0.0000171\\
      9.0  &     329.8859561 &  0.0001041\\
     10.0  &     357.6820849 &  0.0000182\\
     11.0  &     385.4776994 &  0.0000196\\
     12.0  &     413.2735513 &  0.0000183\\
     13.0  &     441.0692677 &  0.0000165\\
     14.0  &     468.8650981 &  0.0000182\\
     15.0  &     496.6609713 & 0.0000188\\
     16.0  &     524.4569189 &  0.0000181\\
     17.0  &     552.2525673 &  0.0000189\\
     18.0  &     580.0482405 &  0.0000184\\
     19.0  &     607.8443134 &  0.0000183\\
     20.0  &     635.6401325 &  0.0000111\\
     22.0  &     691.2315569 &  0.0000183\\
     23.0  &     719.0272818 &  0.0000183\\
     24.0  &     746.8231388 &  0.0000186\\
     25.0  &     774.6189715 &  0.0000200\\
     26.0  &     802.4149260 &  0.0000165\\
     27.0  &     830.2107265 &  0.0000183\\
     28.0  &     858.0062450 &  0.0000164\\
     29.0  &     885.8022693 &  0.0000108\\
     30.0  &     913.5981742 &  0.0000191\\
     32.0  &    969.1896806  & 0.0000185\\
     34.0  &    1024.7813973 &  0.0000180\\
     35.0  &    1052.5769963 &  0.0000181\\
     36.0  &    1080.3729364 &  0.0000185\\
     37.0  &    1108.1688588 &  0.0000178\\
     38.0  &    1135.9643473 &  0.0000171\\
     39.0  &    1163.7602825 &  0.0000165\\
     40.0  &    1191.5562184 &  0.0000165\\
     41.0  &    1219.3520276 &  0.0000170\\
     42.0  &    1247.1477983 &  0.0000172\\
     43.0  &    1274.9435043 &  0.0000174\\
     44.0  &    1302.7392995 &  0.0000176\\
     45.0  &    1330.5350990 &  0.0000165\\
     46.0  &    1358.3310048 &  0.0000162\\
     47.0  &    1386.1268850 &  0.0000162\\
     49.0  &    1441.7183129 &  0.0000170\\
     50.0  &    1469.5143195 &  0.0000169\\
     51.0  &    1497.3100851 &  0.0000167\\\hline
\enddata
\end{deluxetable*}

\begin{deluxetable*}{lll}
\centering
\caption{Secondary eclipse timings of the Kepler-34 binary.}\label{tab:SETVs}
\tablehead{
\colhead{Index} & \colhead{Timing (BJD-2454900)} & \colhead{Uncertainty (day)}}
\startdata
      0.0   &     69.1799238 &  0.0000364\\
      2.0   &    124.7716625 &  0.0000380\\
      3.0   &    152.5673717 &  0.0000373\\
      4.0   &    180.3628899 &  0.0000381\\
      5.0   &    208.1588538 & 0.0000366\\
      6.0   &    235.9543960 &  0.0000360\\
      7.0   &    263.7500467 &  0.0000355\\
      8.0   &    291.5463136 &  0.0000395\\
      9.0   &    319.3417153 &  0.0000363\\
     10.0   &    347.1374054 &  0.0000369\\
     11.0   &    374.9332081 &  0.0000575\\
     12.0   &    402.7288062 &  0.0000396\\
     13.0   &    430.5250276 &  0.0000351\\
     14.0   &    458.3205897 &  0.0000348\\
     15.0   &    486.1163858 &  0.0000368\\
     16.0   &    513.9120140 &  0.0000381\\
     17.0   &    541.7075405 &  0.0000367\\
     18.0   &    569.5037809 &  0.0000096\\
     19.0   &    597.2997462 &  0.0000351\\
     20.0   &    625.0944156 &  0.0000355\\
     22.0   &    680.6865661 &  0.0000358\\
     23.0   &    708.4823252 &  0.0000365\\
     25.0   &    764.0740494 &  0.0000346\\
     26.0   &    791.8693645 &  0.0000376\\
     27.0   &    819.6650814 &  0.0000365\\
     29.0   &    875.2570230 &  0.0000378\\
     30.0   &    903.0492716 &  0.0006748\\
     31.0   &    930.8484250 &  0.0000394\\
     32.0   &    958.6440631 &  0.0000352\\
     33.0   &    986.4397517 &  0.0000380\\
     34.0   &   1014.2358519 &  0.0000343\\
     35.0   &   1042.0314760 &  0.0000390\\
     36.0   &   1069.8272041 &  0.0000358\\
     38.0   &   1125.4187726 &  0.0000358\\
     39.0   &   1153.2145266 &  0.0000350\\
     40.0   &   1181.0103966 &  0.0000360\\
     41.0   &   1208.8060216 &  0.0000365\\
     42.0   &   1236.6016353 &  0.0000369\\
     43.0   &   1264.3973238 &  0.0000373\\
     44.0   &   1292.1935444 &  0.0000376\\
     45.0   &   1319.9890590 &  0.0000355\\
     47.0   &   1375.5805224 &  0.0000349\\
     48.0   &   1403.3761629 &  0.0000346\\
     49.0   &   1431.1722013 &  0.0000403\\
     51.0   &   1486.7634601 &  0.0000359\\
     52.0   &   1514.5597914 &  0.0002017\\\hline
\enddata
\end{deluxetable*}

\end{CJK*}
\end{document}